\documentclass[aps,twocolumn,graphics,floatfix,tightenlines]{revtex4-2}
\usepackage{graphics}
\usepackage{subcaption}
\usepackage{epstopdf}
\usepackage{epsfig}
\usepackage{graphicx}
\usepackage{epsf,epic}
\usepackage{color}
\usepackage{mathtools}
\usepackage{amsmath}
\usepackage{booktabs}
\usepackage{multirow}
\usepackage{physics}
\usepackage{hyperref}
\usepackage{amsfonts}
\usepackage{wrapfig}
\usepackage{breqn}
\usepackage{pstricks}
\usepackage[utf8x]{inputenc}
\usepackage{multirow}
\usepackage{fancyref}
\usepackage{pst-node}
\usepackage{soul}
\usepackage{float}
\usepackage{bm}
\usepackage{dcolumn}
\newcommand{\etal}{\textit{et al.\ }}
\newcommand{\ie}{\textit{i.e.\ }}

\makeatletter
\usepackage{etoolbox} 
\appto{\appendix}{%
	\@ifstar{\def\theequation@prefix{A.}}%
	{}%
}
\preto\maketitle{%
	\begingroup\lccode`~=`,
	\lowercase{\endgroup
		\let\saved@breqn@active@comma~
		\let~}\active@comma 
}
\appto\maketitle{%
	\begingroup\lccode`~=`,
	\lowercase{\endgroup
		\let~}\saved@breqn@active@comma 
}
\makeatother
\begin{document}	
	\title{Altermagnetism of MnSiN$_2$ and MnGeN$_2$ in a wurtzite-based crystal structure}
	\author{Ilteris K. Turan}
	\email{ilteris.turan@case.edu}
	\author{Jonathan Agrinya}
	\author{Walter R L. Lambrecht}
	\email{walter.lambrecht@case.edu}
	\affiliation{Department of Physics, Case Western Reserve University, 10900 Euclid Avenue, Cleveland, OH 44106-7079, USA}
	
	\begin{abstract}
		Based on symmetry analysis the $Pna2_1$ wurtzite-based crystal structure with Mn as magnetic group-II element, MnSiN$_2$ and MnGeN$_2$  belong to the class of altermagnets because the spin up and spin down Mn in the unit cell are related by rotation or mirror plane type operations rather than inversion or translation. Spin polarized density functional calculations and quasiparticle self-consistent (QS) $GW$ calculations support the symmetry analysis and identify the lowest energy spin configuration to be $Pna^\prime2_1^\prime$ in agreement with neutron diffraction data. The altermagnetic band splittings allowed for the three different antiferromagnetic ordering patterns are reported and symmetry analyzed. Additionally, spin wave dispersions, chiral magnon splitting, and N\'eel temperature estimates from the mean field and random phase approximations are provided as well as calculations of the optical dielectric function using the Bethe Salpeter Equation method. MnSiN$_2$ exhibits both Wannier type excitons with N-$p$ to Ge(Si)-$s$
		transition character and charge transfer Frenkel type excitons with hole on N-$p$ and electron on Mn-$d$ orbital character.
	\end{abstract}
	
	\maketitle
	\section{Introduction}
	MnSiN$_2$ and MnGeN$_2$ belong to the family of II-IV-N$_2$ semiconductors exhibiting the wurtzite-based $Pna2_1$ structure (prototype $\beta$-NaFeO$_2$) which is an ordered arrangement of the group-II and group-IV cations on the wurtzite lattice accompanied by structural distortions. They have tetrahedral coordination similar to MnGeP$_2$ in the chalcopyrite $I\bar{4}2d$ spacegroup. The choice of a magnetic group II element Mn with a half-filled $d$-shell results in the formation of a  magnetic moment and as usual in semiconductors, an antiferromagnetic arrangement of the spins with a net zero magnetization. However, if the spin up and spin down Mn are related by (in this case screw) rotation or (glide) mirror operations instead of a translation or inversion, the resulting structure is altermagnetic, a new classification of collinear magnetism that was recently introduced \cite{Mazin2022,Smejkal2022,Smejkal2022a,YuanLD2020}. This is the case in both the chalcopyrite structure and the $Pna2_1$ structure.
	Distinguishing it from usual antiferromagnets, the energy bands are then no longer spin-degenerate at all {\bf k}-points \cite{YuanLD2020,Smejkal2022} and spin waves can be chiral.\cite{Smejkal2023,Sears2026,Beida2025}
	In a recent paper \cite{Turan2025} we showed that MnGeP$_2$ is indeed an altermagnet but in experimental samples realized thus far, this altermagnetism is hidden by the presence of a ferromagnetic secondary phase MnP. In contrast, antiferromagnetism in MnSiN$_2$ and MnGeN$_2$ has already been reported in 1977 and 1972 by Wintenberger \etal \cite{Wintenberger1977,Wintenberger1972}. More recently, below the antiferromagnetic transition a canting transition of the magnetic moments was reported for MnSiN$_2$ \cite{Kautzsch2023}, but the altermagnetic nature has not been recognized until now. Other relevant prior studies on the crystal growth and magnetic ordering include \cite{Esmaeilzadeh2006,Lazarov2004}.
	
	Here we consider both MnSiN$_2$ and MnGeN$_2$ and report first-principles calculations of the electronic band structures using the quasiparticle self-consistent (QS) \cite{MvSQSGWprl,Kotani2007} version of the many-body-perturbation theory (MBPT) $GW$ method where $G$ and $W$ stand for the one-electron Green's function and screened Coulomb interaction as introduced by Hedin \cite{Hedin65,Hedin69}. We use symmetry analysis to explain the directions in {\bf k}-space where spin-degeneracy is maintained and report the spin splittings for low-symmetry directions for each of the three possible antiferromagnetic spin arrangements. We also calculate the total energy differences between ferromagnetic, and the different antiferromagnetic arrangements. Additionally, we use a linear response method to determine the exchange interactions and calculate the N\'eel transition temperatures and spin wave dispersions, and we estimate the amount of chiral magnon splitting (in the absence of dipolar splitting \cite{Sears2026}). Finally, we calculate the lowest energy structure's dielectric function and the optical absorption coefficients, obtained within the independent particle approximation (IPA) and Bethe-Salpeter equation (BSE), and compare with the experimental measurements of \cite{Esmaeilzadeh2006}. 
	
	\section{Methodology and Theoretical Background}
	
	The calculations are performed using the {\sc Questaal} suite of codes \cite{questaalpaper}, which implement both Density Functional Theory (DFT) and the Many-Body Perturbation Theory (MBPT) QS$GW$ \cite{Kotani2007} and Bethe-Salpeter-Equation (BSE) methods \cite{Cunningham21,Cunningham23}
	using a muffin-tin orbital basis set. The exchange interactions
	are calculated using the linear response approach of Kotani and van Schilfgaarde \cite{Kotani2008}.
	
	A well converged basis set of double $\{\kappa,R_{sm}\}$ smoothed Hankel functions, where $\kappa$ controls the decay rate of the functions and $R_{sm}$ controls the smoothing radius which describes the width of the Gaussian functions with which the Hankel functions are convoluted \cite{questaalpaper} in the envelope function. These envelope functions are constructed with angular momentum cut-offs of $spdfspd$ on all atoms and augmented in the spheres up to $\ell_{max}=4$. Mn $4d$ orbitals are added as local orbitals to improve the accuracy of conduction band states. The numerical values of the complete basis sets used can be found at \cite{datavail}.
	
	The QS$GW$ in the RPA is a variant of the $GW$ method, in which a non-local  exchange correlation potential $\tilde{\Sigma}_{xc}$ is extracted from the dynamic self-energy, $\Sigma(\omega)$, via
	\begin{equation}
		\tilde{\Sigma}_{xc}=\frac{1}{2}{\rm Re}\left\{\sum_{nm}|\psi_n\rangle[\Sigma_{nm}(\epsilon_n)+\Sigma_{nm}(\epsilon_m)]\langle\psi_m|\right\},
		\label{eqvxc}
	\end{equation}
	where Re stands for the taking the Hermitian part of the following terms, and $\epsilon_n$, $|\psi_n\rangle$ are the eigenvalues and eigenstates of the effective one-particle Hamiltonian $H^0$. The full procedure is as follows: from a given DFT level $H^0$'s eigenstates and eigenvalues, the one particle Green's function $G^0$ and the polarization propagator  $P^0=-iG^0G^0$ are calculated and provide the screened Coulomb interaction $W^0=(1-P^0v)^{-1}v$ and following Hedin's formalism \cite{Hedin65, Hedin69}, the self-energy convolution integral $\Sigma(\omega)=iG^0W^0$ is calculated as a matrix in the basis of $H^0$ eigenstates. Then $\tilde{\Sigma}_{xc}$ is extracted from Eq.(\ref{eqvxc}) and its difference from the original DFT $v_{xc}$ is added to $H^0$ for the next iteration until the self-energy no longer changes within a given tolerance. This scheme, in principle, makes the results independent of the DFT starting point as long as the latter provided a reasonable approximation to the correct eigenstates. 
	
	A $4\times4\times4$, $\Gamma$-centered {\bf k}-mesh is used for charge density and self-consistency sampling of the Brillouin zone and a $3\times3\times3$ mesh is used to sample the $GW$ dynamical self energy $\Sigma({\bf k})$. At each iteration of QS$GW$, the charge density is made self-consistent, and the tolerance value is chosen to be $10^{-5}$ Ry. The self-energy matrix is approximated by an average diagonal energy above 27.21 eV (2 Ry) for MnSiN$_2$ and above 35.65 eV (2.62 Ry) for MnGeN$_2$.
	
	Typically, the band gaps are somewhat overestimated in QS$GW$ in the random phase approximation (RPA) because of the underestimate of the screened Coulomb interaction $W({\bf q},\omega)$. To mitigate this error, we used a recent version of QS$GW$, named QS$GW^{\textrm{BSE}}$, which goes beyond the RPA by including electron-hole interactions in $W({\bf q},\omega)$ or ladder diagrams in the polarization propagator via a BSE approach \cite{Cunningham23}. The correction to the band gap is sensitive to the number of bands used in the BSE treatment of $W$. Here we employ this method for the lowest energy $Pna^\prime2_1^\prime$ structures, with a set of top 26 valence bands and bottom 13 conduction bands to be included in the BSE kernel, which covers the majority of the bands with Mn-$d$ and N-$p$ characters. A similar BSE approach can be used to calculate the optical matrix elements in the long wavelength ${\bf q}\rightarrow 0$ limit, which captures the exciton peaks in the macroscopic dielectric function \cite{Onida02,Strinati1988,Hanke78}.
	
	In the IPA, we find the imaginary part of the optical dielectric function $\varepsilon_2$, by calculating the velocity matrix elements for direct transitions from the valence states $\psi_{v{\bf k}}$ to conduction states $\psi_{c{\bf k}}$ with energies $\epsilon_{v{\bf k}}$, $\epsilon_{c{\bf k}}$ in $GW^{\textrm{BSE}}$:
	
	\begin{equation}
		\begin {split}
			\varepsilon_2(\omega)&=\frac{8\pi^2}{\Omega \omega^2} \sum_{cv{\bf k}}(f_{v{\bf k}}-f_{c{\bf k}})\left|\bra{\psi_{v{\bf k}}}\hat{\bm{e}}\cdot \bm{v}\ket{\psi_{c{\bf k}}}\right|^2 \\ &\times\delta(\epsilon_{c{\bf k}}-\epsilon_{v{\bf k}}-\omega),
		\end{split}
	\label{eps2}
	\end{equation}
	where $\Omega$ is the unit cell volume, $\hat{\bm{e}}$ the polarization, $\bm{v}$ the velocity operator, and $f_{n{\bf k}}$ are Fermi occupation factors for the conduction bands $c$ and valence bands $v$.
	
	In the BSE in the Tamm-Dancoff approximation \cite{Hanke78,Strinati1988,Rohlfing2000,Onida02}, the dielectric function is similarly calculated from the eigenvalues and eigenvectors of a two-particle Hamiltonian,
	\begin{eqnarray}
	H^{(2p)}_{n_1n_2{\bf k},n_1^\prime n_2^\prime{\bf k}^\prime}({\bf q})&=&\left(\epsilon_{n_2{\bf k}+{\bf q}}-\epsilon_{n_1{\bf k}}\right)\delta_{n_1n_1^\prime}\delta_{n_2n_2^\prime}\delta_{{\bf kk}^\prime}\nonumber\\
	&&-\left(f_{n_2{\bf k}+{\bf q}}-f_{n_1{\bf k}}\right) K_{n_1n_2{\bf k},n_1^\prime n_2^\prime{\bf k}^\prime}({\bf q})\nonumber, \\
	\end{eqnarray}
	with $f_{n{\bf k}}$ the Fermi occupation function for band $n$ at ${\bf k}$,
	which includes the electron-hole interactions kernel $K_{n_1n_2{\bf k},n_1^\prime n_2^\prime{\bf k}^\prime}({\bf q})$ expanded in the basis of one-particle eigenstates $\psi_{n\bf k}(\bf r)$. This effectively mixes the direct vertical band to band transitions $(\epsilon_{c{\bf k}}-\epsilon_{v{\bf k}})$ at different {\bf k} into new ``exciton'' eigenstates $E^\lambda$. 
	Diagonalizing this Hamiltonian in the Tamm-Dancoff approximation, where $n_1$ is restricted to be a valence state and $n_2$ a conduction band state,
	one obtains the exciton eigenvalues $E^\lambda({\bf q})$ and eigenvectors
	$A^\lambda_{n_1n_2{\bf k}}({\bf q})$, \ie $A^\lambda_{vc{\bf k}}({\bf q})$. Here $\lambda$ denotes the excitonic eigenstate. The macroscopic dielectric function is then given by,
	\begin{eqnarray}
	\varepsilon_M(\omega)&=&1-\lim_{{\bf q}\rightarrow0}\frac{8\pi}{|{\bf q}|^2\Omega N_k}
	\sum_{ss^\prime}(f_{n_2^\prime{\bf k}^\prime+{\bf q}}-f_{n_1^\prime{\bf k}^\prime})\nonumber \\
	&&\rho_s({\bf q}) \sum_\lambda\frac{A^\lambda_s({\bf q})A^{\lambda*}_s({\bf q})}{E^\lambda({\bf q})-\omega\pm i\eta} \rho_{s^\prime}({\bf q})^*,
	\label{epsmac}
	\end{eqnarray}
	where we introduced the shorthand $s=\{n_1n_2{\bf k}\}$, with the matrix elements,
	\begin{equation}
	\rho_{n_1n_2{\bf k}}({\bf q})=\langle\psi_{n_2{\bf k}+{\bf q}}|e^{i{\bf q}\cdot{\bf r}}|\psi_{n_1{\bf k}}\rangle.
	\end{equation}
	Here, a factor two is included for spin and $N_k$ is the number of {\bf k}-points in the Brillouin zone. We note that neither of these methods include indirect band to band transitions assisted by the electron-phonon coupling, therefore we expect the absorption to occur only after the direct band gap from $\Gamma$ to $\Gamma$ in the IPA.

	As for the magnetic properties, the Heisenberg exchange interactions are calculated using the linear response approach of Kotani and van Schilfgaarde \cite{Kotani2008}. Briefly, in this method the spin waves are obtained as the poles of the transverse susceptibility defined as $\chi^{+-}({\bf r,r'},t-t')=-i\langle T[\hat{S}^{+}({\bf r},t)\hat{S}^{-}({\bf r}',t')]\rangle_{\text{gs}}$  where $\langle ... \rangle_{\text{gs}}$ denotes the expectation value for the ground state, $T(...)$ the time ordering operator, and $\hat{S}^{\pm}=\hat{S}^x \pm i\hat{S}^y$ the spin raising and lowering operators. Its Fourier transform for the bare (non-interacting) version of this response function is calculated from the energy bands and wave functions using
 \begin{equation}
		\begin{split}
			\chi_{\bm{q}}^{0+-}(\bm{r}, \bm{r}',\omega)&=\\\sum_{\bm{k}n\downarrow}^{occ.}\sum_{\bm{k}'n'\uparrow}^{unocc.} \frac{\psi^{*}_{\bm{k}n\downarrow}(\bm{r}) \psi_{\bm{k}'n'\uparrow}(\bm{r})\psi^{*}_{\bm{k}'n'\uparrow}(\bm{r}')\psi_{\bm{k}n\downarrow}(\bm{r}')}{\omega-(\epsilon_{\bm{k}'n'\uparrow}-\epsilon_{\bm{k}n\downarrow})+i\delta}\\
			+\sum_{\bm{k}n\downarrow}^{unocc.}\sum_{\bm{k}'n'\uparrow}^{occ.}\frac{\psi^{*}_{\bm{k}n\downarrow}(\bm{r}) \psi_{\bm{k}'n'\uparrow}(\bm{r})\psi^{*}_{\bm{k}'n'\uparrow}(\bm{r}')\psi_{\bm{k}n\downarrow}(\bm{r}')}{-\omega-(\epsilon_{\bm{k}n\downarrow}-\epsilon_{\bm{k}'n'\uparrow})+i\delta},\label{eq:chi}
		\end{split}	
	\end{equation} 
where $\bm{k}'=\bm{k}+\bm{q}$, and $\psi_{{\bf k}n\sigma}$ and $\epsilon_{{\bf k}n\sigma}$ are the spin-dependent band structure eigenstates and eigenvalues, which in our case are calculated in the QS$GW$ approximation.       
        Using a rigid spin approximation (RSA) we then integrate the propagator $\chi^{0+-}({\bf q},\omega)$ over the volume $\Omega_a$ of each atomic spheres $a$ using a projector $\bar{e}_a(\bf{r})$ proportional to the local magnetization density, $\bar{e}_a({\bf r})=m_a({\bf r})m_a/\int_{\Omega_a} m_a({\bf r})^2d^3r$ 
	\begin{equation}
		D^0_{aa'}({\bf q},\omega)=\int_{\Omega_a} d^3r\int_{\Omega_a'}d^3r' \bar{e}_a({\bf r})\chi^{0+-}({\bf q},\omega)\bar{e}_{a'}({\bf r}'). 
	\end{equation}
	The inverse of this matrix
	$[{\bf D}^{0+-}({\bf q},\omega=0)]^{-1}={\bf J}({\bf q})$ is then identified with the matrix of exchange interactions between pairs of atoms in the unit cell, and is related to the real space exchange interactions by a Bloch sum or discrete Fourier transform $J^c_{ab}({\bf q})=\sum_{\bf T}e^{i{\bf q}\cdot{\bf T}} J^c_{a,b+{\bf T}}$, where $\textbf{T}$ are lattice translation vectors. These pertain to the exchange interactions of the classical Heisenberg Hamiltonian
	\begin{equation}
		H=-\sum_{a,b,{\bf T},{\bf T}^\prime} J^c_{a+{\bf T},b+{\bf T}^\prime}\bm{e}_{a{\bf T}}\cdot\bm{e}_{b{\bf T}^\prime},
		\label{eqHeis}
	\end{equation}
	in which the size of the magnetic moments is absorbed in the $J^c$, thus leaving $\bm{e}_{a{\bf T}}$ unit vectors. Alternatively, we can also calculate the dressed transverse susceptibility assuming an on-site only interaction kernel chosen so as to obey the sum-rules. The poles of this dynamic susceptibility coincide with the spin waves and their broadening reflects the life time of the spin waves. This quantity is visualized as a heat-map of the imaginary part of the trace of the matrix. 
	
	From the real space exchange interactions, one obtains the mean field approximation (MFA) critical temperature by constructing the $J^0_{ab}=J^c_{ab}({\bf q}=0)=\sum_{\bf T}J_{a,b+{\bf T}}$ omitting the $J^c_{aa}$ on-site term for ${\bf T}=0$. The highest eigenvalue of this matrix $j_{\text{max}}$ gives $T_c^{\text{MFA}}=\frac{2}{3k_B}j_{\text{max}}$ for the case of multiple magnetic atoms per cell. The MFA typically gives an overestimate of the actual $T_c$. 
	
	We also use the random phase approximation (RPA) of Tyablikov and Callen \cite{Tyablikov1959,Callen63,Rusz2005}. In this method, one constructs the matrix $\textbf{N}$, with elements
	\begin{equation}
		N_{ab}({\bf q})=-\langle S_a\rangle \bar{J}_{ab}({\bf q}),
		\label{eqNmatrix}
	\end{equation}
	with,
	\begin{equation}
		\bar{J}_{ab}=J_{ab}({\bf q}) -\sum_c \langle S_a\rangle^{-1}J_{ac}({\bf q}=0)\langle S_c\rangle
	\end{equation}
	where $\langle S_a\rangle $ is the thermal average of the spin on sublattice $a$  and the $J_{ab}=J^c_{ab}/|S_aS_b|$ are exchange interactions of the Heisenberg Hamiltonian written in terms of spins $S_a$. In section \ref{sec:sw}, we discuss that the eigenvalues of a similar form of the $N$-matrix yields the spin wave energies. In the RPA method, one then calculates the Brillouin zone average of the 
	inverse of the $N$-matrix, and uses the consistency equation 
	\begin{equation}
		\langle S_a \rangle=\frac{2}{3}\beta_c\left\{\int_{BZ} d^3q [N^{-1}({\bf q};\langle S_a\rangle)]_{aa}\right\}^{-1}, \label{eqIa}
	\end{equation}
	giving the new spin average on each sublattice in terms of $N$ constructed from the same spin average to determine $\beta_c=1/k_BT_c$.
	Because at $T_c$ the average spin vector becomes scale independent, we use a normalized vector and $\beta_c$ is the scaling factor between the new and old normalized vectors ${\langle S_a \rangle}$, giving an estimate of $T_c$ at each iteration, 
	\begin{equation}
		T_c^{RPA}=\frac{2}{3/k_B}\frac{1}{I_a\langle S_a\rangle},
	\end{equation}
	where $I_a$ is the Brillouin zone integral in Eq.\ref{eqIa}.
	The normalized eigenvector is initiated by the mean field eigenvector of ${\bf J}({\bf 0})$ matrix and is iterated until the normalized $\langle S_a \rangle_{\rm new}$ given by the left hand side of Eq.\ref{eqIa} (without the factor $\beta_c$) equals the $\langle S_a\rangle_{\rm old}$ on the right hand side in constructing $N$ within tolerance, \ie at the fixed point of the consistency equation. We notice that if only one type of magnetic atom is present, the spin vector is just determined by the signs of the spins in the reference configuration, and the initial $T_c$ are already close to the converged $T_c$. The scheme then does not take many iterations to converge. In fact, if the process yields other $T_c$ values, this occurs at locations where the $\langle S_a \rangle$ vectors evolve to a different fixed point with a different spin configurations, where they either change sign or drastically change in magnitude compared to the initialized vector. We check that each sub-lattice gives the same $T_c$ and that the final $\langle S_a\rangle$ are close to the initial one from the MFA and correspond to the spin arrangement in the reference configuration. 

	\begin{figure*}[t]
		\includegraphics[width=18cm]{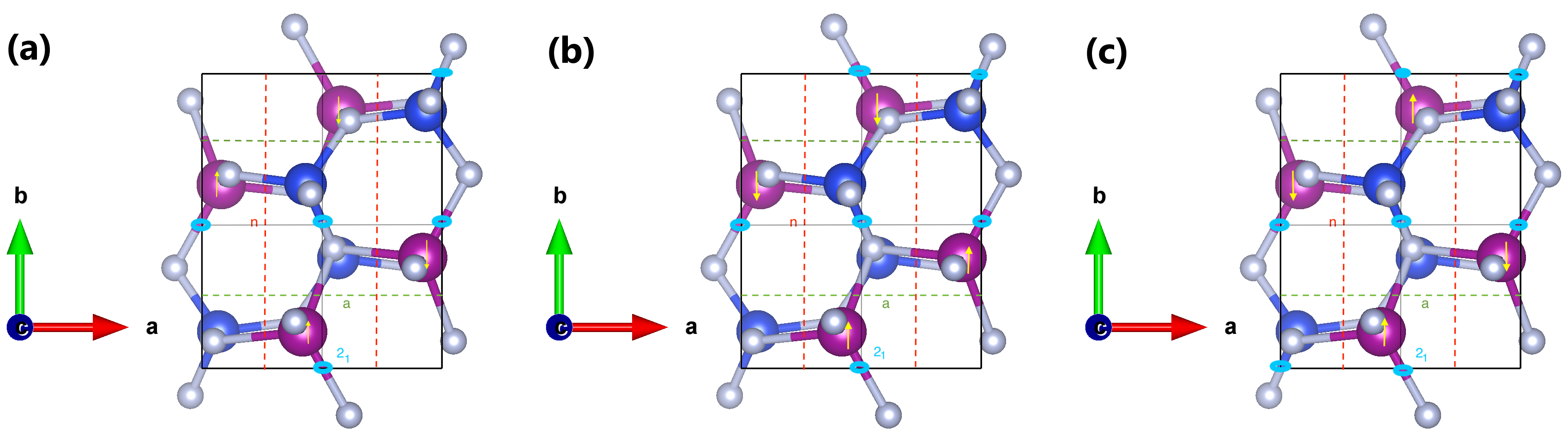}{}
		\caption{Crystal structure and magnetic ordering of Mn atoms; magenta Mn, blue Si or Ge, grey N; glide mirror plane $n$ (red dashed line) includes glides by $b/2,c/2$, glide mirror plane $a$ (dashed green line) includes glide by $a/2$, turquoise ovals $2_1$ screw axis along $z$; yellow arrows indicate spin; (a) $Pna^\prime2_1^\prime$, (b) $Pn^\prime a2_1^\prime$, (c) $Pn^\prime a^\prime 2_1$.}\label{fig.magorder} 
	\end{figure*}	
			
	The difficulty is that $\textbf{N}({\bf q}=0)$ diverges. To avoid the divergence one calculates instead the Green's function ${\bf }G(z)=[z{\bf I}+{\bf N}({\bf q}=0)]^{-1}$ for some points $z$ in the complex plane with positive imaginary part and uses analytic continuation towards the real axis at $z\rightarrow0$.\cite{Rusz2005,Hass1984} The distance of the points in the complex plane is set by a parameter $\epsilon$ and the adequacy of the approximation is guided by the requirement that the imaginary part of the inverse is small compared to the real part and that for points close to {\bf q}=0, the analytic continuation and the direct evaluation of the inverse coincide. The RPA method requires a fine ${\bf q}$-mesh to adequately sample the region near ${\bf q}\rightarrow 0$.
	The four-point analytic continuation
	$f(0)=4f(i\epsilon)-f(2i\epsilon)-f(\epsilon+i\epsilon)-f(-\epsilon+i\epsilon)$ suggested by Rusz \etal \cite{Rusz2005} is used for each of the sublattice $G_{aa}(z)$.

	\begin{figure*}[t]
		\includegraphics[width=18cm]{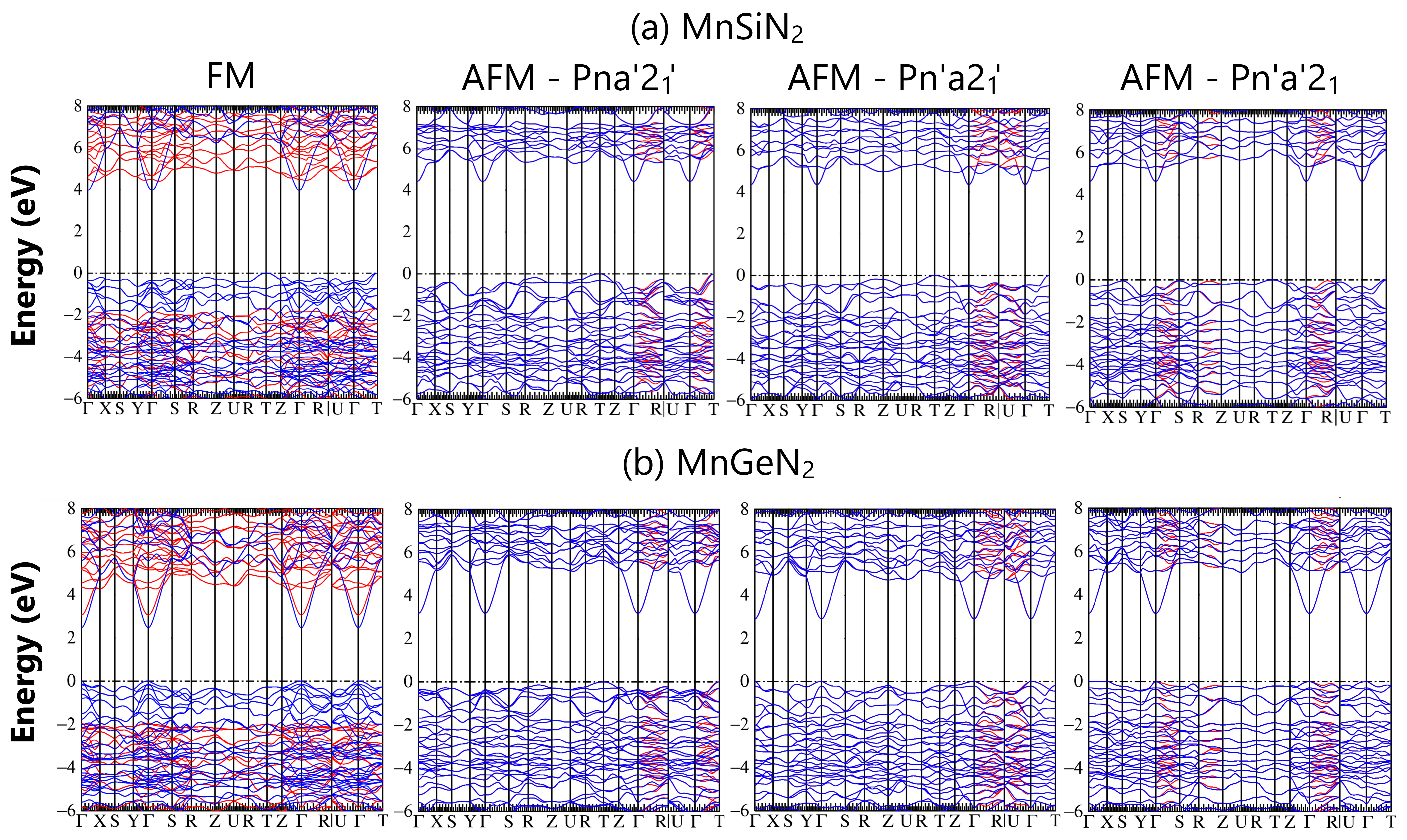}{}
		\caption{Energy bands of in QS$GW$ approximation for (a) MnSiN$_2$ and (b) MnGeN$_2$ plotted in order: FM $Pna2_1$, and AFM $Pna^\prime2_1^\prime$, $Pn^\prime a 2_1^\prime$, $Pn^\prime a^\prime 2_1$. The spin up bands are shown in blue and spin down bands are shown in red.} \label{fig:bnds} 
	\end{figure*}	
	
	\section{Results}
	\subsection{Magnetic orderings and total energy differences}
	
	In Fig. \ref{fig.magorder} we show the three different magnetic configurations of the Mn atoms in the crystal structure. The symmetry operations are indicated and those that require a spin flip are indicated by a prime in the magnetic spacegroup notation. Besides these three antiferromagnetic arrangements, we also consider the ferromagnetic arrangement. The total energy differences with respect to the lowest energy
	$Pna^\prime2_1^\prime$ are given in Table \ref{tab:totalenergy} as obtained in spin-polarized density functional theory using the PBEsol exchange correlation functional.
	We note that the lowest energy configuration is in agreement with the neutron diffraction study of Wintenberger \etal \cite{Wintenberger1977} and Esmaeilzadeh \etal \cite{Esmaeilzadeh2006} but in disagreement with Kautzsch \etal \cite{Kautzsch2023} who claimed the ordering to be
	$Pn^\prime a 2_1^\prime$. In fact, Wintenberger \etal also denoted the magnetic group as $Pn^\prime a2_1^\prime$ in the text but their figure shows it is  $Pna^\prime2_1^\prime$.
	All studies agree that the $2_1$ rotation must be accompanied by a spin-flip. As our calculation here does not include spin-orbit coupling it does not tell us about the orientation of the spin relative to the crystal axes, only the relative orientation of the spins on different Mn atoms with respect to each other. 
	
	\begin{table}[h]
		\caption{Total energy differences of different magnetic configurations.} \label{tab:totalenergy}
		\begin{ruledtabular}
			\begin{tabular}{lcc}
				\multicolumn{3}{c}{MnSiN$_2$} \\ \hline
				&spacegroup & $\Delta E$ (meV/atom) \\ \hline
				AFM &    $Pna^\prime2_1^\prime$   & 0 \\
				AFM &    $Pn^\prime a^\prime 2_1$ & 31.9 \\
				AFM &    $Pn^\prime a2_1^\prime$  & 48.2 \\
				FM  &    $Pna2_1$                 & 90.6 \\ \hline
				\multicolumn{3}{c}{MnGeN$_2$} \\ \hline
				AFM &    $Pna^\prime2_1^\prime$   & 0 \\
				AFM &    $Pn^\prime a^\prime 2_1$ & 27.4 \\
				AFM &    $Pn^\prime a2_1^\prime$  & 42.4 \\
				FM  &    $Pna2_1$                 & 76.5\\
			\end{tabular}
		\end{ruledtabular}
	\end{table}	
	
	\subsection{Energy bands and symmetry analysis}
	
	\begin{figure}
		\includegraphics[width=7cm]{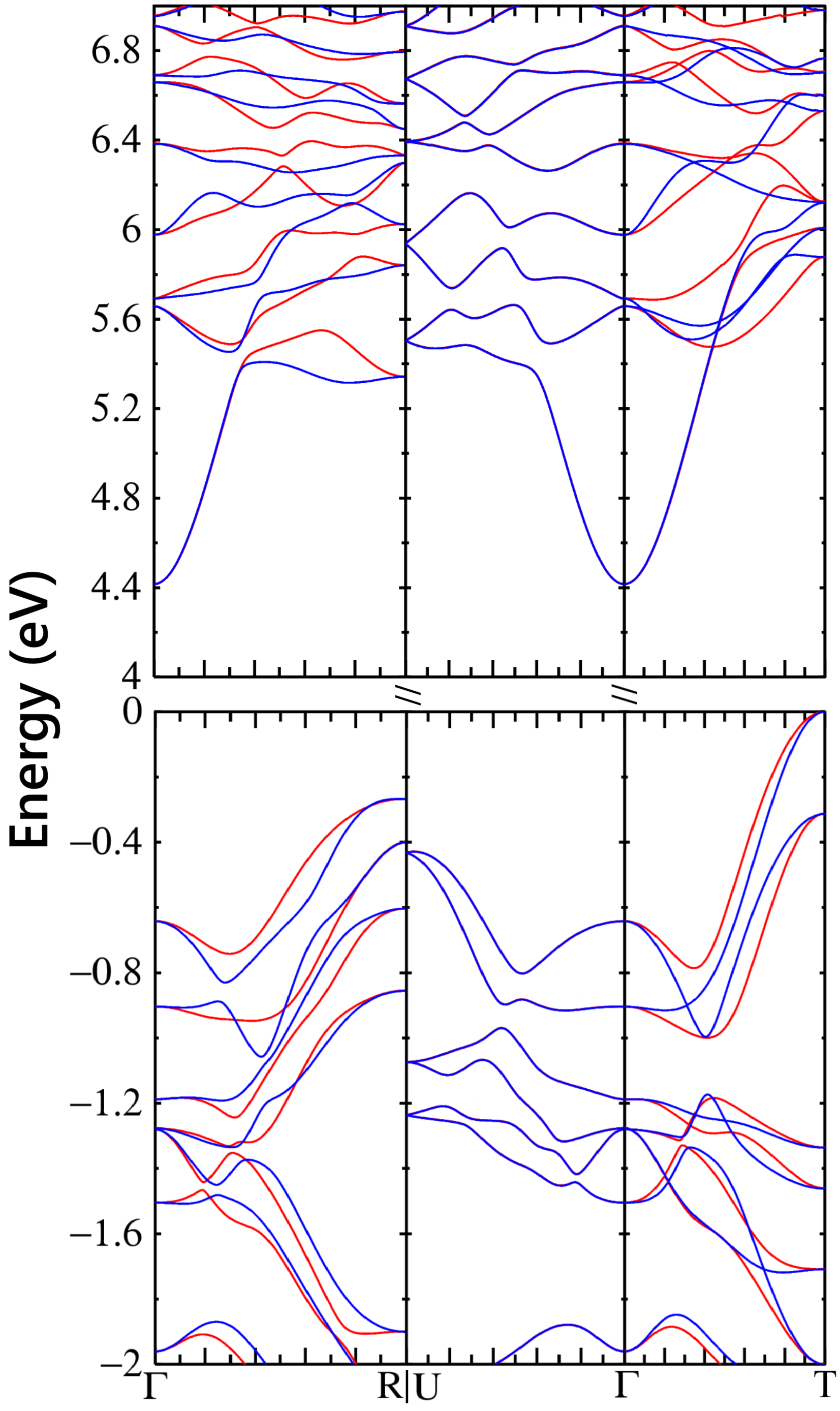}{}
		\caption{Zoom in on VBM and CBM of the energy bands
			of $Pna^\prime2_1^\prime$ MnSiN$_2$. Majority spin bands are colored blue and minority spin red. The maximum amount of splitting in the valence bands along $\Gamma R$ is 120 meV and along $\Gamma T$ is 150 meV.} \label{fig:bndzoom}
	\end{figure}
	
	In an altermagnet, the energy bands at arbitrary {\bf k} are not spin degenerate. However for {\bf k} for which the little group of {\bf k} contains at least one element in common with the point group part of the spin spacegroup operation that flips the spin, there is still a degeneracy. In addition, we need to consider time reversal symmetry because there is no net magnetization. The irreducible part of the orthorhombic Brillouin zone is a rectangular box with corners $\Gamma$ $(0,0,0)$, $X$ $(\pi/a,0,0)$, $Y$ $(0,\pi/b,0)$, $S$ $(\pi/a,\pi/b,0)$, $Z$ $(0,0,\pi/c)$, $U$ $(\pi/a,0,\pi/c)$, $T$ $(0,\pi/b,\pi/c)$, $R$ $(\pi/a,\pi/b,\pi/c)$.  For any point in the $\Gamma XYS$ plane the two-fold rotation turns ${\bf k}$ into $-{\bf k}$
	which are related by time reversal, which turns spin up into spin down. Thus for all these points, the bands remain spin-degenerate if the spin group contains $2_1^\prime$.
	Likewise, for any point in the $ZURT$ plane $k_x\rightarrow -k_x$, $k_y\rightarrow -k_y$ and $k_z=\pi/c$ stays fixed under the $C_{2z}$ operation but is equivalent to $k_z=-\pi/c$ by the reciprocal lattice vector $(0,0,2\pi/c)$ and hence the transformed {\bf k} is equivalent to $-{\bf k}$ which means the spatial part of the wave function goes into its complex conjugate. Together with the spin reversal associated with $2_1^\prime$ this means the bands in the $ZURT$ plane stay degenerate if the magnetic group contains $2_1^\prime$ by time reversal symmetry.
	Furthermore, for $Pna^\prime2_1^\prime$ any point in the
	$\Gamma ZXU$ plane lies in the mirror plane $m_y$ perpendicular to ${\bf b}$, so these points turn into themselves under this mirror operation which also flips the spin. Having this symmetry operation in common between the group of {\bf k}  and the set $C_{2z}\{E,m_x\}$,
	the band must stay spin-degenerate in this plane. Here $\{E,m_x\}$ is the subgroup of point operations, corresponding to identity and the $n$ glide mirror plane of the space group which leaves the crystal invariant without spin flips and $C_{2z}$ the two fold rotation about the $z$ axis which together with $C_{2z}m_x=m_y$ corresponding to the $a$-glide mirror in real space
	constitute the spin flip requiring operations of the spin group. Thus, for example along $\Gamma U$ the bands should stay spin-degenerate but along $\Gamma T$ they can split for $Pna^\prime2_1^\prime$.
	Any of the high-symmetry lines bounding the irreducible Brillouin zone lies in at least one plane where the spin-degeneracy holds. However, 
	for a general direction such as $\Gamma R$ which has no symmetry at all, the  bands can split except at the end points.  Similar considerations apply for the other antiferromagnetic configurations. One may also note that for points in the $XSUR$ plane the group of {\bf k} contains $m_x$ because this operation brings it to a point related by a reciprocal lattice vector. Thus if we have $n^\prime$ in the magnetic spacegroup the bands anywhere in $XSUR$ remain spin degenerate. Likewise if we have $a^\prime$, bands along $YSTR$ are spin-degenerate.  
	The {\bf k} along which we expect spin-degeneracy to hold for the three antiferromagnetic structures are summarized in Table \ref{tab-sym}.
	
	\begin{table}
		\caption{List of {\bf k} planes which remain spin-degenerate.}\label{tab-sym}
		\begin{ruledtabular}
			\begin{tabular}{ccccccc}
				& $\Gamma XSY$ & $\Gamma XUZ$ & $\Gamma YTZ$ & $ZURT$ &$XSUR$ & $YSTR$ \\ \hline
				$Pna^\prime2_1^\prime$ &  yes  & yes & no  & yes  & no  & yes    \\
				$Pn^\prime a2_1^\prime$ &  yes  & no  & yes  & yes  & yes  & no     \\
				$Pn^\prime a^\prime2_1$ &  no  & yes  & yes & no   & yes  & yes    \\
			\end{tabular}
		\end{ruledtabular}
	\end{table}
	
	\begin{table}[h]
		\caption{Band gaps of MnSi(Ge)N$_2$ in QS$GW$ (eV). The values in parenthesis for  $Pna^\prime2_1^\prime$ are the band gaps obtained within QS$GW^{\textrm{BSE}}$. Both band gaps for the FM configurations are from majority spin bands.}\label{tab:gaps}
		\begin{ruledtabular}
			\begin{tabular}{lcc}
				\multicolumn{3}{c}{MnSiN$_2$} \\ \hline
				spin group & indirect ($T \rightarrow \Gamma$) &direct ($\Gamma \rightarrow \Gamma$)\\ \hline
				$Pna^\prime2_1^\prime$  & 4.418 (4.151) & 5.058 (4.793)\\
				$Pn^\prime a2_1^\prime$ & 4.355 & 4.826 \\
				$Pn^\prime a^\prime 2_1$ & 4.640 & 4.966 \\
				FM ($\uparrow \uparrow$)  $Pna2_1$ & 3.965 & 4.352 \\ \hline
				\multicolumn{3}{c}{MnGeN$_2$} \\ \hline
				$Pna^\prime2_1^\prime$  & 3.178 (2.983) & 3.544 (3.359) \\
				$Pn^\prime a2_1^\prime$ & 3.102 & 2.905 \\
				$Pn^\prime a^\prime 2_1$ & 3.345 & 3.147 \\
				FM ($\uparrow \uparrow$)   $Pna2_1$ & 2.751 & 2.479\\
			\end{tabular}
		\end{ruledtabular}
	\end{table}
	
	The energy bands of the FM and each of the three AFM configurations are shown in Fig. \ref{fig:bnds} confirming the symmetry analysis of the {\bf k} lines along which spin degeneracy is maintained. For example for $Pna^\prime 2_1^\prime$ we can see the spin degeneracies along the $\Gamma XSY\Gamma S RZURTZ$ and $\Gamma U$ but not along  $\Gamma R$, $\Gamma T$ where spin splittings occur. The band structures of $Pn^\prime a2_1^\prime$ also maintains the spin degeneracy along  $\Gamma XSY\Gamma S RZURTZ$  but now the bands along $\Gamma T $ are degenerate but not along $\Gamma U$. For $Pn^\prime a^\prime 2_1$  the bands show spin splitting along $\Gamma S $ and $ZR$. In the FM case, we see a large splitting between majority and minority spin bands with the top valence bands having majority spin and the low lying conduction bands mostly minority spin but the lowest gap is nonetheless between majority spin electrons at $\Gamma$.
	
	The band gaps are summarized in Table \ref{tab:gaps}.
	The band gap is seen to be indirect with the valence band maximum (VBM) at $T$ and the conduction band minimum at $\Gamma$ in all cases but in $Pn^\prime a^\prime 2_1$ the VBM at $S$ and in between $RZ$ are only slightly lower. The lowest direct gap occurs at $\Gamma$. However, this is not the case in the PBEsol calculation for MnSiN$_2$. This is because in the QS$GW$ calculation, not only is the gap increased, it also shifts up the minority spin Mn-$3d$ states higher than the Si-$s$ like
	CBM at $\Gamma$. In MnGeN$_2$, the deeper Ge-$s$ levels guarantee a conduction band minimum at $\Gamma$ already in the PBEsol DFT.  
	The $\Gamma T$ indirect gap in PBEsol is 1.587 eV and the direct gap at $\Gamma$ is 2.280 eV. Thus the self-energy correction to the gap is about 2.85$\pm0.02$ eV. 
	Typically, this is overestimated by $\sim20$ \% in QS$GW^{RPA}$
	because of the underestimate of the screening.  This would reduce the gaps by about 0.6 eV. Thus we estimate an indirect gap of 3.8 eV and direct gap of 4.4 eV for the $Pna^\prime2_1^\prime$ configuration. Using the QS$GW^{BSE}$ given in parentheses for the $Pna^\prime2_1^\prime$ case, but with a limited number of bands included we see that the gap decrease is somewhat smaller, $\sim$0.2--0.3 eV.
	
	A zoom in on the VBM and CBM for $Pna^\prime2_1^\prime$ MnSiN$_2$ is shown in Fig. \ref{fig:bndzoom}. It shows that the altermagnetic spin splittings along major $\Gamma R$ and $\Gamma T$ lines is of order 0.1 eV (approximately 120 meV and 150 meV) for the top valence bands but only occurs at about 0.8 eV below the VBM near $\Gamma$. For MnGeN$_2$, the valence band splitting is of the same order of magnitude, which reaches to a maximum of approximately 180 meV along $\Gamma T$. Likewise in the conduction band it only occurs at about 1 eV above the CBM. Unfortunately, this limits their effects on electronic transport properties. Only for p-type doped material, one may expect to see the effects of the spin splitting of the constant energy surface for energies just below the VBM.	
	
    \begin{figure}
		\includegraphics[width=8cm]{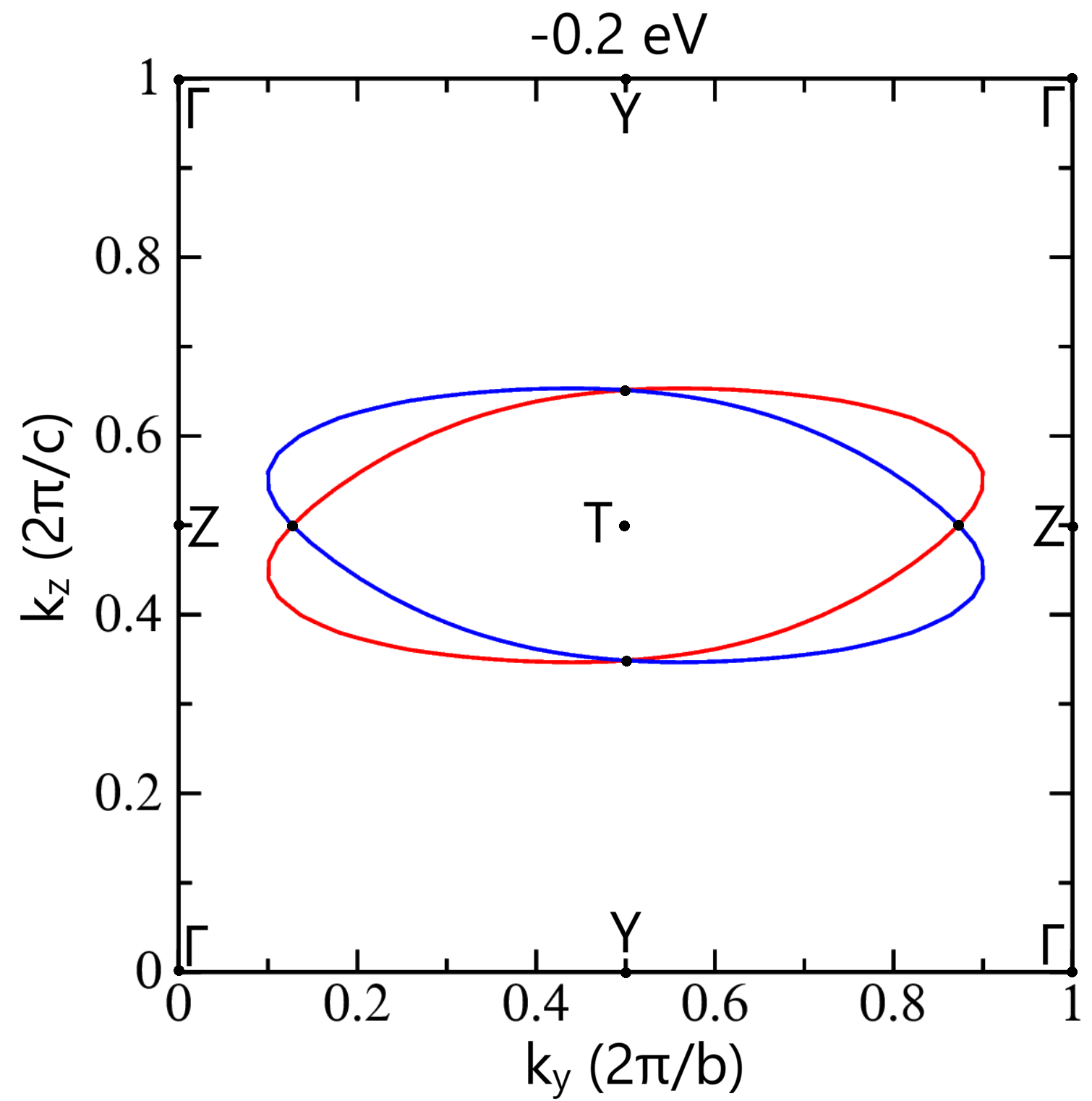}{}
		\caption{Valence energy bands of $Pna'2_1'$ MnGeN$_2$ is shown as a contour plot indicating $d$-wave altermagnetism. A planar cut is taken at 0.2 eV below the Fermi level on the $\Gamma YTZ$ plane where $T$ is centered. The axis labels are parameterized as $[0,1]$ to fit within a total distance from $\Gamma$ to $\Gamma$ between the two neighboring unitcells along $k_y$ (horizontal) and along $k_z$ (vertical) lines.} \label{fig:contour}
	\end{figure}
	
    In Fig. \ref{fig:contour}, we show a contour plot of the constant energy surface in  the $\Gamma YTZ$ plane 0.2 eV below the Fermi energy in $Pna'2_1'$ MnGeN$_2$. As expected, we see  the cross-section of an ellipsoidal energy surface  but which is rotated from the crystal axes in opposite direction for spin up and spin down bands.
    The effective mass tensor was calculated at point $T$ by fitting an ellipsoidal energy surface to a random set of {\bf k}-points near $T$. It gives as principal values  $m_x=3.725\mp0.012$, $m_y= 0.811\pm0.012$ and $m_z=0.465\pm0.12$
    for spin up and down respectively and with principal axes perpendicular to the chosen plane or in the $k_x$ direction toward $R$, $\pm12^\circ$ and  $\pm78^\circ$ from the $TZ$ or $y$-direction in descending order of the principal values. The very high mass along $TR$  and equal by symmetry for up and down spin indicates an almost cylindrical energy surface.
    For orthorhombic symmetry, one would expect the mass tensor to have principal axes along the crystal directions. This is indeed the case for the $x$ direction  which corresponds to the mirror plane $n_x$ which stays a valid symmetry operation without spin reversal. But the $y$ and $z$ directions correspond to $a_y^\prime$ and $2_1^\prime$ which require a spin flip. The average of spin up and spin down energy contours indeed would be an ellipse with principal axes along $TZ$ and $TY$ but the individual spin constant energy surfaces are rotated from these directions so that the $TZ$ and $TY$ directions are the nodal directions.
	Fig. \ref{fig:contour} confirms that the spin splitting reverses sign 
    between $T \Gamma$ and the orthogonal direction $T \Gamma'$ towards the next unit cell along $[01-1]$ and the pattern with two nodal lines means that we can classify the material as a $d$-wave altermagnet in terms of its Fermi surface for slightly p-type doped material.
				
	\subsection{Magnetic moments}
	
	\begin{figure}
		\includegraphics[width=8cm]{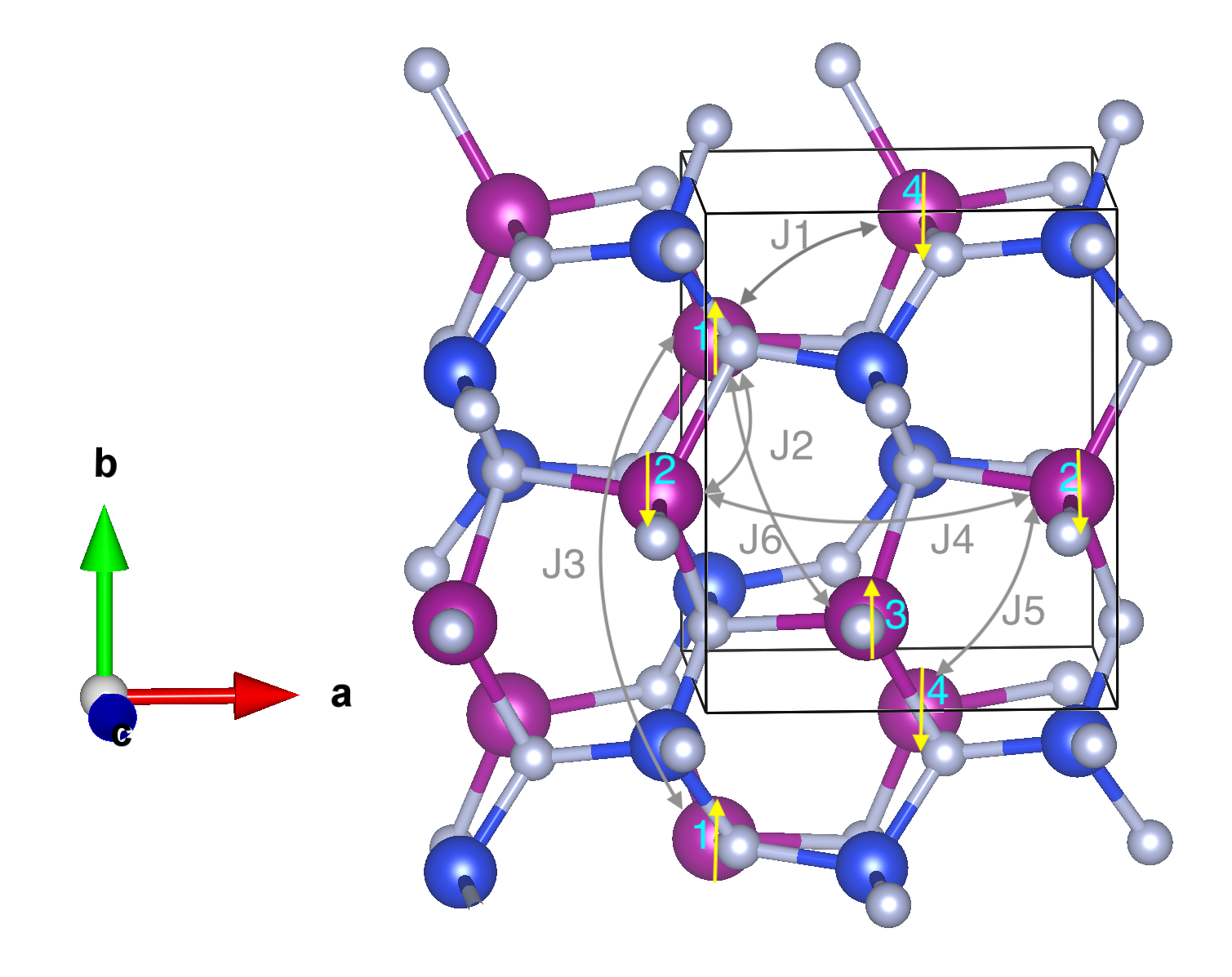}
		\caption{Atom labeling in unit cell and exchange interactions. Table \ref{tab:J} lists the magnitudes of the exchange interactions extracted from $(\chi^{0+-})^{-1}$.}\label{fig:J}
	\end{figure}
	The magnetic moments per Mn are given in Table \ref{tab:mom}. The Mn moments are close to the expected value for a half-filled $d$ shell spin $S=5/2$ and are seen to be largest in the ferromagnetic case. The moments induced on N are small but larger than on the Si or Ge atoms, consistent with N being the nearest neighbors of the magnetic Mn atoms and their moments spilling over to the N. For both of the materials, antiferromagnetic $Pna'2_1'$ structure yield the smallest moment on the alternating spin up and spin down Mn compared to the other antiferromagnetic and ferromagnetic structures. The ferromagnetic structures have a total magnetic moment of 20 $\mu_B$ per unit cell, whereas all of the antiferromagnetic cases yields 0 $\mu_B$ at the end of the convergence cycles. 
	
	\begin{table*}
		\caption{Magnetic moment magnitudes (in $\mu_B$) in each atomic sphere for each magnetic configuration. Values in parenthesis are extrapolated beyond the sphere. N$_1$ indicates the N above Si (Ge), and N$_2$ above Mn.} \label{tab:mom} 
		\begin{ruledtabular}
			\begin{tabular}{ccccc}
				\multicolumn{5}{c}{MnSiN$_2$} \\ \hline
				structure            & Mn            & Si            &   N$_1$       & N$_2$ \\ \hline
				$Pna^\prime2_1^\prime$ & 4.549 (4.737) & 0.001  & 0.005 (0.007) & 0.011 (0.015) \\
				$Pn^\prime a2_1^\prime$ & 4.596 (4.817) & 0.0016 & 0.025 (0.064) & 0.013 (0.017) \\
				$Pn^\prime a^\prime2_1$ & 4.580 (4.791) & 0.0006 & 0.011 (0.015) & 0.023 (0.053) \\
				FM  $Pna2_1$          & 4.695 (4.889) & 0.0268 & 0.035 (0.082) & 0.036 (0.072) \\ \hline
				\multicolumn{5}{c}{MnGeN$_2$} \\ \hline 
				& Mn            & Ge            &   N$_1$       & N$_2$ \\ \hline
				$Pna^\prime2_1^\prime$ & 4.500 (4.746) & 0.0011 (0.0032)  & 0.002 & 0.0014 (0.0059) \\
				$Pn^\prime a2_1^\prime$ & 4.539 (4.811) & 0.002  & 0.025 & 0.0013 (0.0037) \\
				$Pn^\prime a^\prime2_1$ & 4.526 (4.791) & 0.0003 (0.0006) & 0.0032 (0.0034) & 0.021 \\
				FM  $Pna2_1$          & 4.568 (4.867) & 0.035 & 0.034 & 0.035  \\ 
			\end{tabular}
		\end{ruledtabular}
	\end{table*}
	
	\subsection{Exchange interactions and N\'eel temperature}
	
	\begin{figure*}
		\includegraphics[width=18cm]{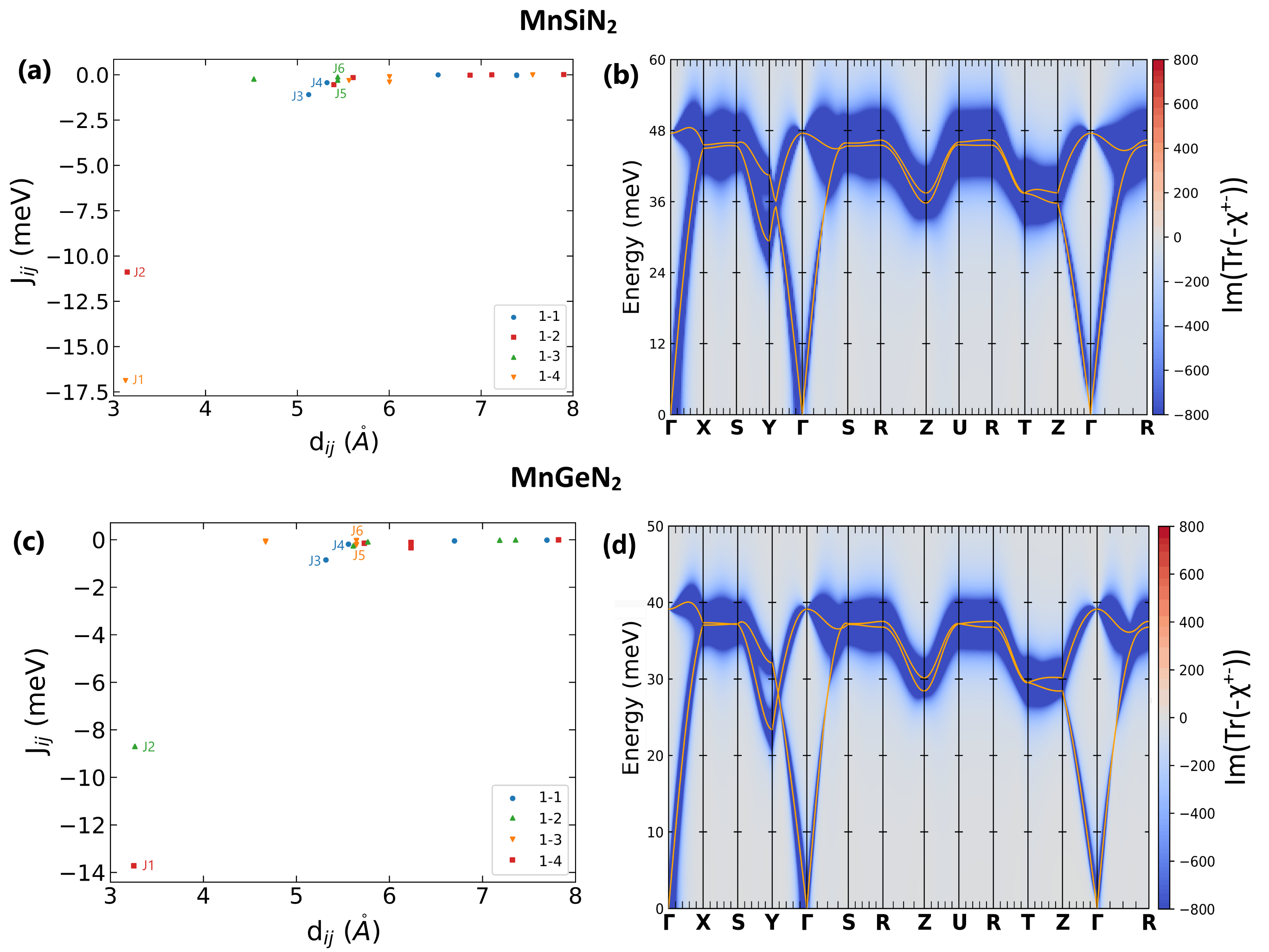}
		\caption{Exchange interactions $J_{1j}$ in meV are shown as function of distance $d_{1j}$ in \AA\  for (a) MnSiN$_2$, (c) MnGeN$_2$; the values are given in table \ref{tab:J} and the labeling of $J$s are shown in Fig. \ref{fig:J}. Spin wave dispersions of (b) MnSiN$_2$, (d) MnGeN$_2$ are shown where the orange line is found from the eigenvalues of the ${\bf N}({\bf q})$ matrix \ref{eqNmatrix}, and the heat-map background shows the imaginary part of the trace of the dynamic $-\chi^{+-}({\bf q},\omega)$. The peaks of $\chi^{+-}({\bf q},\omega)$ align with the spin wave energies, and the maximums occur along $\Gamma X$ at $\sim50$ meV and $\sim40$ meV respectively.} \label{fig:J-d-sw}			
	\end{figure*}	
	
	The main exchange interactions in the $Pna^\prime2_1^\prime$  are given in Table \ref{tab:J} as indicated in Fig. \ref{fig:J} and shown as function of distance for each pair of Mn atoms in the unit cell in Fig. \ref{fig:J-d-sw}.
	We can see that  the largest antiferromagnetic (\ie negative) exchange interaction is between atoms 1 and 4 in the cell which are linked by the $a^\prime$ operation followed by that between 1 and 2 which are linked by the $2_1^\prime$ interaction and thus indeed coupled antiferromagnetically. The interaction between atoms 1 and 3, and 1 and 1 in neighboring cells are also antiferromagnetic but much weaker.
	The strong antiferromagnetic interaction between atoms 1 and 4, and 1 and 2 is already found even if we start from the ferromagnetic reference state and thus can be used to predict the antiferromagnetic ordering consistent with the total energy results. Additionally, the exchange interactions between Mn and the other nonmagnetic Ge(Si), N$_1$, and N$_2$, as well as their interactions among themselves are completely negligible, and these interactions are not included in spin wave and critical temperature calculations. 
	
	The mean field N\'eel temperature obtained from these exchange interactions is 423 K for MnSiN$_2$ and 339 K for MnGeN$_2$. For MnSiN$_2$, this is remarkably close to the experimental values of 443 K \cite{Kautzsch2023}, 453 K \cite{Wintenberger1977}, 442 K \cite{Esmaeilzadeh2006} but this may be in part due to cancellation of errors. For a spin 5/2 system the quantum  correction $(S+1)/S=1.4$ which would bring the estimate to 592 K but the mean field approximation tends to overestimate critical temperatures by a similar factor.
	For MnGeN$_2$ the experimental values are found to be only slightly lower than for MnSiN$_2$ at 448 K \cite{Wintenberger1972}. Using the Random Phase Approximation \cite{Tyablikov1959,Callen63,Rusz2005} we find for MnSiN$_2$ 230$\pm10$ K when treating the moments as classical vectors and using a $50\times50\times50$ {\bf q}-mesh and applying the analytic continuation approach of Rusz \etal \cite{Rusz2005}
	to avoid the divergence of $[N^{-1}({\bf q})]_{aa}$ at ${\bf q}=0$ with an $\epsilon$ of about 5 K. At that value we obtain the same result as omitting the point ${\bf q}=0$ and we obtain a ratio of the imaginary to the real part of the corresponding Green's function of order 10$^{-4}$. Doubling the number of {\bf q}-point divisions, yielded essentially the same result, showing adequate convergence. 
	After multiplying by  the quantum correction factor for the overall spin of $S=5/2$, in other words, a factor $7/5=1.4$ we find 322 K, which gives a lower limit of the critical temperature. The average value of MFA and RPA estimates scaled by the quantum correction factor, yields critical temperature of 457 K, which is in agreement with the experimentally observed values.
	
	\begin{table}
		\caption{Exchange interactions $J_{ij}({\bf d})$ in MnSi(Ge)N$_2$. The vector {\bf d} connecting atoms $i$ and $j$ is given in Cartesian coordinates and d=$|{\bf d}|$ in \AA.} \label{tab:J}.
		\begin{ruledtabular}
			\begin{tabular}{cccccc}
				\multicolumn{6}{c}{MnSiN$_2$} \\ \hline
				site& site & {\bf d} & d &name & $J$ (meV) \\ \hline
				1   &   4  &  (2.661,-1.647,0.000) & 3.130 & $J_1$ & -16.94 \\
				1   &   2  &  (0.855, 1.619, 2.561) & 3.148 & $J_2$ & -10.94 \\
				1   &   1  &  (0,0,5.123) & 5.123 & $J_3$ & -0.92 \\
				1   &   1  &  (5.322,0,0) & 5.322 & $J_4$ & -0.31 \\
				1   &   3  & (-1.806,3.266,-2.561) & 4.526 & $J_5$ & -0.25 \\
				1   &   4  &   (-1.806,3.266,2.561) &4.526 & $J_6$ &  -0.24 \\ \hline
				\multicolumn{6}{c}{MnGeN$_2$} \\ \hline
				1   &   4  &  (2.780,-1.688,0.000) & 3.252 & $J_1$ & -13.72 \\
				1   &   2  &  (0.905, 1.662, 2.758) & 3.345 & $J_2$ & -8.70 \\
				1   &   1  &  (0,0,5.317) & 5.317 & $J_3$ & -0.85\\
				1   &   1  &  (5.560,0,0) & 5.560 & $J_4$ & -0.19 \\
				1   &   3  & (-1.874,3.350,-2.658) & 4.669& $J_5$ & -0.09 \\
				1   &   4  & (-1.874,3.350,2.658) &4.669 & $J_6$ &  -0.06 \\ \hline
			\end{tabular}
		\end{ruledtabular}
	\end{table}
	\begin{figure*} [t]
		\includegraphics[width=18cm]{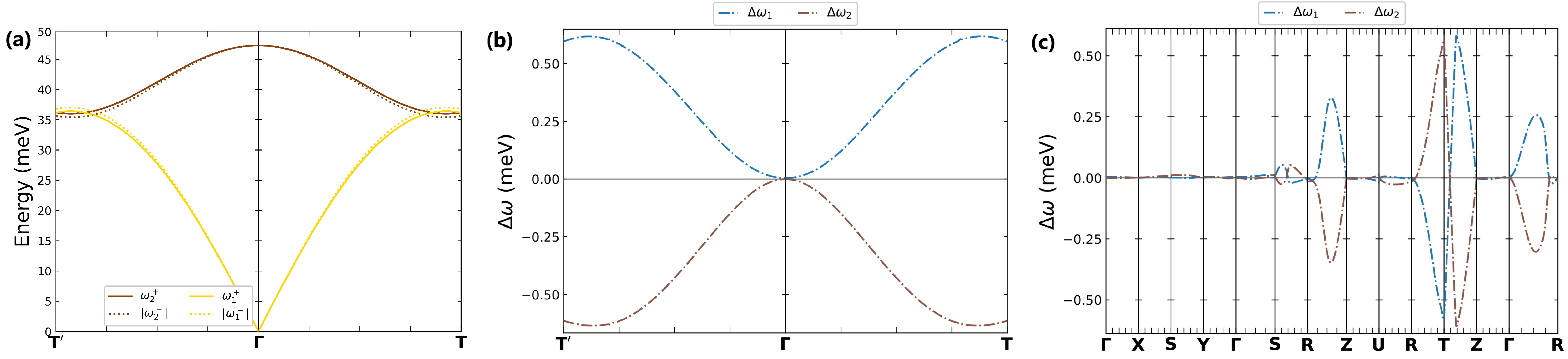}
		\caption{Chiral magnon splitting in $Pna'2_1'$ MnSiN$_2$: (a) Spin wave dispersion of positive (solid) and the absolute value of the negative (dotted) branches in meV along $T'\Gamma T$. The $\omega_2$, shown in brown color, is the higher energy spin wave above the Goldstone mode, $\omega_1$ shown in yellow. The difference of positive and negative spin waves $\Delta \omega_i = \omega_i^{+}-|\omega_i^{-}|$ for $i=1,2$ in meV are shown along (b) $T'\Gamma T$ and (c) full symmetry line. The maximum splitting for both $\omega_1$ and $\omega_2$ is around 0.57 meV.} \label{fig:chiral}
	\end{figure*}

	\subsection{Spin wave dispersions}
	\label{sec:sw}
	The spin-wave dispersion and dynamic ${\rm Im Tr} [-\chi^{+-}({\bf q},\omega)]$ for antiferromagnetic $Pna^\prime2_1^\prime$ structures are shown in Fig. \ref{fig:J-d-sw}(b),(d).
	The peaks in  ${\rm Im Tr} [-\chi^{+-}({\bf q},\omega)]$ correspond to the poles located at the spin wave eigenvalues. The solid lines are found within a linear response approach to the Heisenberg Hamiltonian, where the spin wave energies $\omega(q)$ of both crystals are obtained from the eigenvalue equation,
	\begin{equation}
		\det{\omega({\bf q})\delta_{ab}\frac{1}{2S_a}-\frac{1}{S_a}\tilde J_{ab}^{\cal H}({\bf q})\frac{1}{S_b}}=0\label{eqsw}
	\end{equation}
	with,
	\begin{equation}
		\tilde J_{ab}^{\cal H}({\bf q})=J_{ab}^{\cal H}({\bf q})-\delta_{ab}\frac{1}{S_a}\left(\sum_kJ_{ac}^{\cal H}({\bf q}=0)S_c\right),
	\end{equation}
	where $2S_i=M_i$ for $i=a,b$, $M_i$ are the magnetic moments per sphere and the $J^{\cal H}_{ab}=J^c_{ab}/|S_aS_b|$ are the exchange interactions with respect to spins instead of unit vectors. Note that if we multiply Eq.(\ref{eqsw})
        by $S_a$ it is the same as the $N_{ab}$ matrix 
        used for the RPA estimate of the critical temperature, but here the $S_a$ are the spin values per sublattice in the ordered ground state, rather than the thermal averaged components of the spin. 
	
	Because we have four Mn atoms carrying a magnetic moment, we obtain four distinct spin wave branches, with two positive ($\omega_1^{+}$, $\omega_2^{+}$) and two negative frequencies ($\omega_1^{-}$, $\omega_2^{-}$). For both positive and negative frequencies, the lower energy branch is the acoustic magnon, which corresponds to in-phase precession of the spin sublattices which vanishes at ${\bf q}=(0,0,0)$ (\ie a Goldstone mode), whereas the higher lying optical magnons correspond to the out-of-phase collective precession between the sublattices.  Fig.\ref{fig:chiral}(a) shows the spin wave dispersion along $T'\Gamma T$, where $T'$ lies in the opposite $[0-1-1]$ direction, which includes both the positive $\omega_1^{+}$, $\omega_2^{+}$ and the absolute value of the negative branches $|\omega_1^{-}|$, $|\omega_2^{-}|$. In an antiferromagnet, the negative modes should be equal and opposite to the positive modes for the entire Brillouin zone. However, in an altermagnet the positive and negative branches corresponding to right- and left-handed magnons (\ie the counter-clockwise vs. clockwise precession around the crystal's net magnetization axis) do not need to be each other's mirror image along certain wave vector directions, which is known as the chiral symmetry breaking \cite{Smejkal2023, Sears2026, Beida2025}. In the absence of the long range dipolar interactions and relativistic effects, our calculations predict such splitting to take its maximum value of $\pm 0.57$ meV for MnSiN$_2$ shown in \ref{fig:chiral}(b)(c) and $\pm 0.51$ meV for MnGeN$_2$ near the $T$ point. The other notable $q$-directions which yield large splittings are towards $Z(0,0,1)$ and $R(1,1,1)$ points, which are of sizes $\pm 0.33$ meV and $\pm 0.25$ meV respectively in MnSiN$_2$. 
	 		
	\subsection{Optical properties}
	
	\begin{figure*}
		\includegraphics[width=18cm]{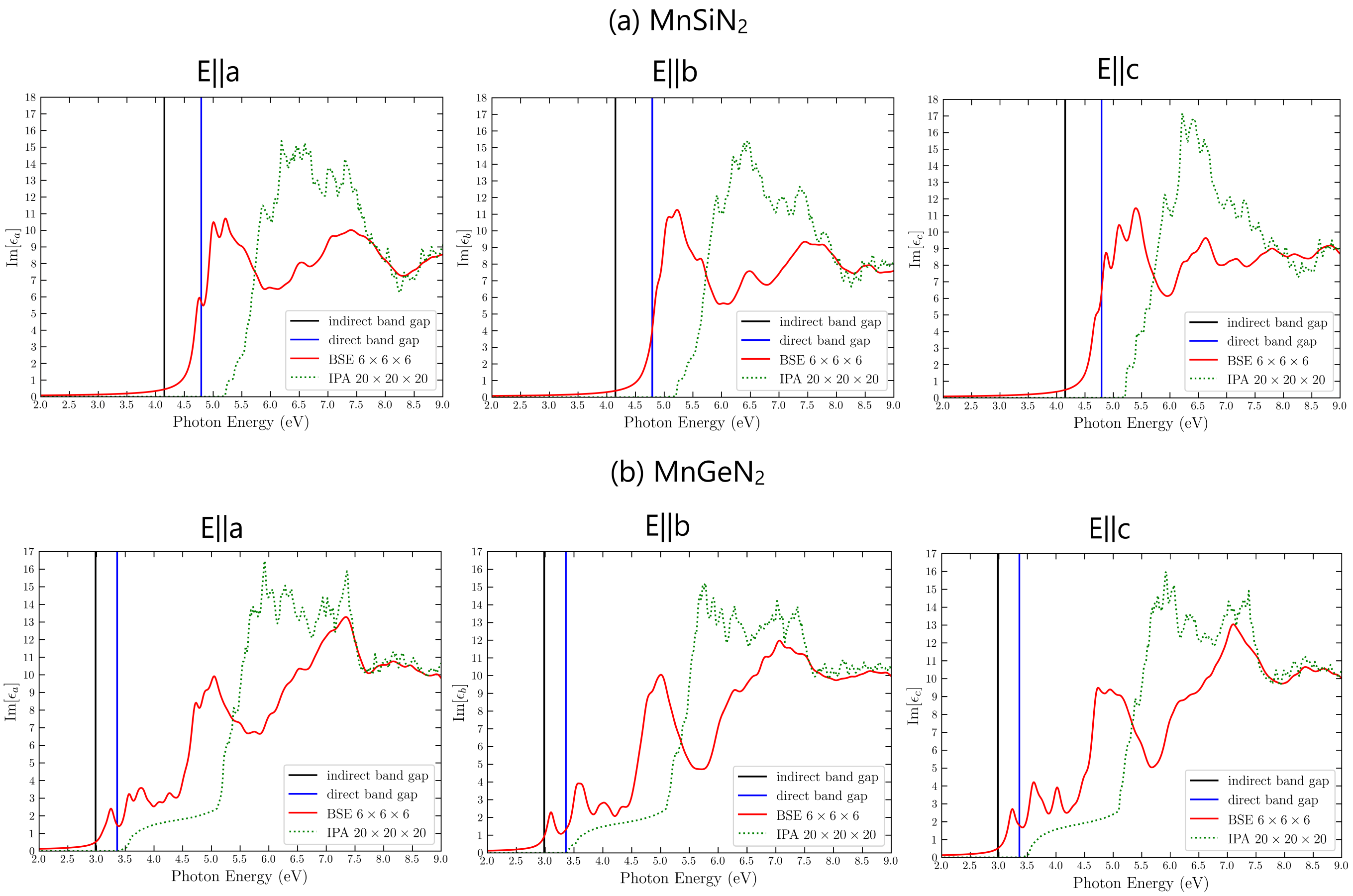}
		\caption{The three components of the imaginary part of the dielectric function parallel to {\bf a}, {\bf b}, {\bf c} axes are shown respectively for (a) MnSiN$_2$ and (b) MnGeN$_2$. The data obtained using the BSE approach are shown in red solid lines. It uses a $6\times6\times6$ {\bf k}-point mesh with a broadening parameter of $\eta=0.005$ Ry (as in equation \ref{epsmac}). The two-particle kernel includes transition matrix elements from 6 valence bands and 3 conduction bands. Independent particle approximation (IPA) results using a finer $20\times20\times20$ {\bf k}-point mesh, without any broadening factor (as in the equation \ref{eps2}) are shown in green dotted lines. The vertical black lines indicate the location of the indirect gap from $T$ to $\Gamma$ within the QS$GW^{\textrm{BSE}}$, whereas the vertical blue lines are that of the smallest direct gap from $\Gamma$ to $\Gamma$.
		\label{fig-eps}}
	\end{figure*}
	\begin{figure*}
		\includegraphics[width=18cm]{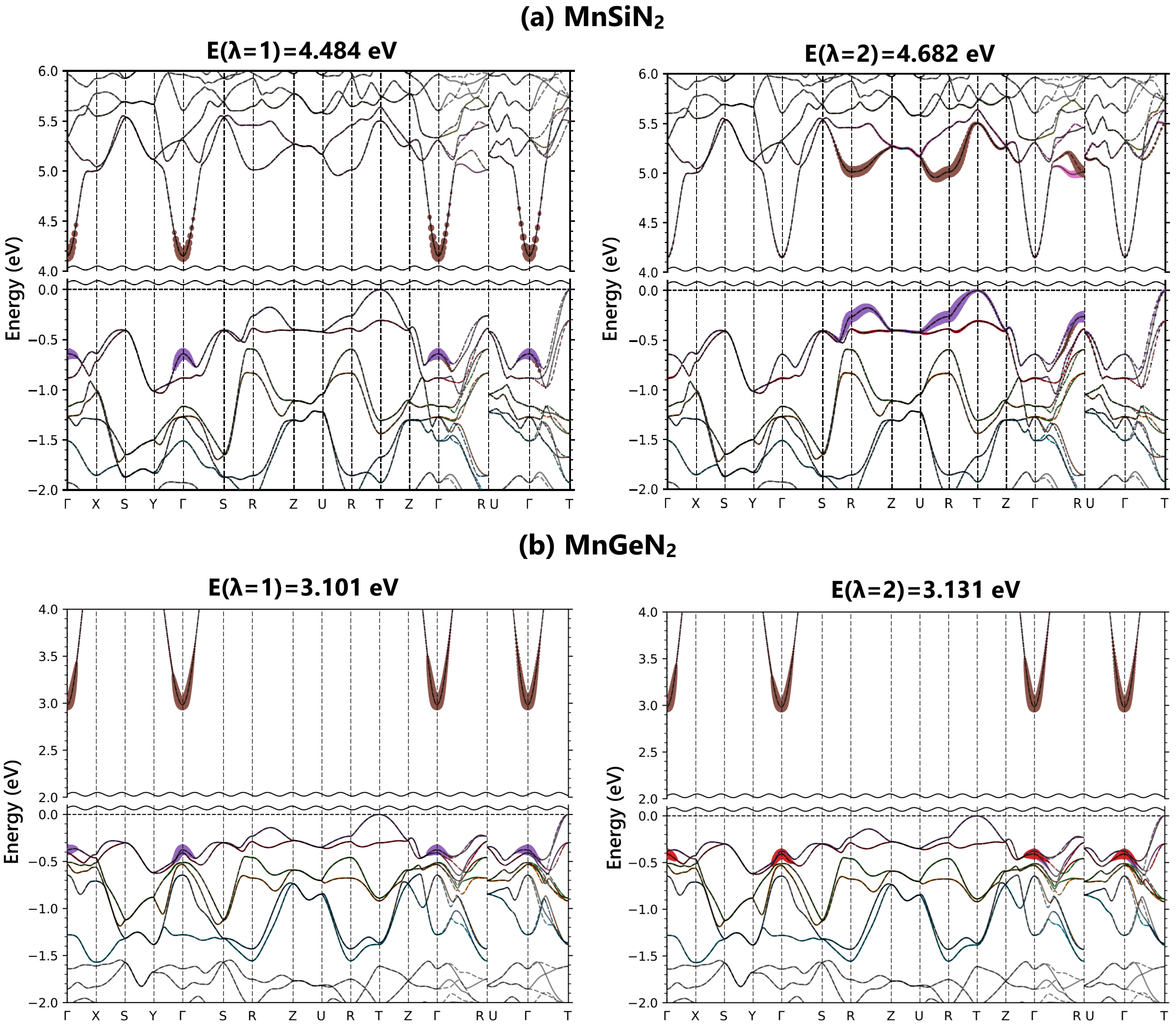}
		\caption{Exciton wave function weights $W_{v(c)\bf k}^{\lambda=1,2}$ in (a) MnSiN$_2$ where $\lambda=1$ exciton from $\Gamma \rightarrow \Gamma$ is Wannier-like, $\lambda=2$ exciton is Frenkel-like, spread out at the top of the Brillouin zone, and in (b) MnGeN$_2$ contributed by CBM of $e$ symmetry and top two valence bands of $e$ and $o$ symmetries. The size of the colored circles are scaled with respect to the exciton weights, and the use of different colors serves to distinguish different bands. Here, the spin-up bands are shown in solid lines and spin down bands are shown in dashed-lines. 
			\label{fig-exciton}}
	\end{figure*}
	
	\begin{table}
		\caption{Symmetry labeled energy levels at $\Gamma$ relative to VBM for the $Pna'2_1'$ structures within QS$GW^{\rm BSE}$ listing the three lowest conduction bands and top six valence bands. The column labels $e$ or $o$ stand for even or odd with respect to the $m_x$ mirror plane.} 
		\label{tablevels}
		\begin{ruledtabular}
			\begin{tabular}{ccc}
				\multicolumn{3}{c}{MnSiN$_2$} \\ \hline
				\multicolumn{1}{c}{level} & \multicolumn{1}{c}{symmetry} & \multicolumn{1}{c}{energy (eV)} \\ \hline
				CBM+2 & $e$ & 5.9756 \\  
				CBM+1 & $o$ &  5.9497 \\
				CBM & $e$ &  4.7927 \\
				VBM & $o$ &  0 \\
				VBM-1 & $e$ & -0.2399 \\
				VBM-2 & $e$ &  -0.5245 \\
				VBM-3 & $e$ &  -0.6270 \\ 
				VBM-4 & $o$ & -0.6303 \\
				VBM-5 & $o$ & -0.8685 \\ \hline 
				\multicolumn{3}{c}{MnGeN$_2$} \\ \hline
				CBM+2 & $o$ &  5.7895 \\  
				CBM+1 & $e$ &  5.6953 \\
				CBM & $e$ &  3.3587 \\
				VBM & $e$ &  0 \\
				VBM-1 & $o$ &  -0.0321\\
				VBM-2 & $e$ &  -0.1364 \\
				VBM-3 & $o$ &  -0.1546 \\ 
				VBM-4 & $e$ &  -0.2691 \\
				VBM-5 & $o$ &  -0.9026 \\  
			\end{tabular}
		\end{ruledtabular}
	\end{table}
 	
 	\begin{table*}
 		\caption{Table shows a list of excitons with band origins, exciton eigenvalues ($E^\lambda$), oscillator strengths ($S^\lambda$) relative to that of the first bright one, polarizations ($\hat e$) if bright, symmetries of the corresponding valence bands (even or odd with respect to $m_x$) at the $\Gamma$ point, the band energies ($\epsilon_c-\epsilon_v$), and exciton binding energies ($E_B$). All results are obtained from the $6\times6\times6$-mesh BSE calculation including matrix elements between six valence and three conduction bands and energies are given in eV.}
 		\label{tab-exciton}
 		\begin{ruledtabular}
 			\begin{tabular}{lccccccc}
 				\multicolumn{8}{c}{MnSiN$_2$} \\ \hline
 				$\lambda$&origin&$E^\lambda$&$S^\lambda$&$\hat e$&sym. at $\Gamma$&$\epsilon_c-\epsilon_v$&$E_B$\\ \hline
 				1&VBM($\Gamma$)$\rightarrow$CBM($\Gamma$)&4.484&0 &dark &$o$ &4.793 &0.309  \\
 				2&VBM($TZUR$)$\rightarrow$CBM($TZUR$)&4.682&1.00 &$z$ &$e$ & &  \\
 				3&VBM$-$1($\Gamma$)$\rightarrow$CBM($\Gamma$)&4.712&0.08 &$y$ and $z$ &$e$ &5.033 &0.321 \\
 				4&VBM and VBM$-$1($TZUR$)$\rightarrow$CBM($TZUR$)&4.747&1.54 &$x$ &$o$ & &  \\
 				5&VBM($\Gamma$)$\rightarrow$CBM($\Gamma$)&4.847&0.01 &$x$ &$o$ &4.793& -0.054\\ \hline
 				\multicolumn{8}{c}{MnGeN$_2$} \\ \hline
 				1&VBM($\Gamma$)$\rightarrow$CBM($\Gamma$)&3.101&1.00 &$y$ &$e$ &3.359 &0.258 \\
 				2&VBM$-$1($\Gamma$)$\rightarrow$CBM($\Gamma$)&3.131&0.19 &$x$ &$o$ &3.391 &0.260 \\
 				3&VBM$-$2($\Gamma$)$\rightarrow$CBM($\Gamma$)&3.233&1.12 &$z$ &$e$ &3.495 &0.262 \\
 				4&VBM$-$3($\Gamma$)$\rightarrow$CBM($\Gamma$)&3.251&0.95 &$x$ &$o$ &3.513 &0.262 \\
 				5&VBM$-$4($\Gamma$)$\rightarrow$CBM($\Gamma$)&3.362&0.25 &$y$ and $z$ &$e$ &3.628 &0.266 \\ 
 			\end{tabular}	
 		\end{ruledtabular}
 	\end{table*}
 	
	The optical dielectric function of MnSiN$_2$ and MnGeN$_2$ are calculated both in the independent particle approximation (IPA) and in the BSE approach for the AFM $Pna^\prime2_1^\prime$ configurations that are based on the QS$GW^{\textrm{BSE}}$ band structures. The IPA method does not include local field nor electron-hole interaction effects while the BSE does include both. The results for the imaginary part of the dielectric function $\varepsilon_2(\omega)$ are shown in Fig. \ref{fig-eps}. In this figure, the IPA results are obtained using a large $20\times20\times20$ {\bf k}-mesh to show clearly the onset at the direct gap. The BSE results, on the other hand, use a coarser $6\times6\times6$ mesh. We can see a clear red-shift of the spectrum and the emergence of bound excitons below the direct gap onset. The dielectric function looks rather different in MnSiN$_2$ and MnGeN$_2$: in MnGeN$_2$ we see a flat region above the gap onset between 3.5 eV and about 5 eV, which is related to the lowest conduction band's strong dispersion in this region. Above this energy new bands come into play. In MnSiN$_2$ the lowest conduction band near $\Gamma$ only lies below the set of rather flat bands related to the empty Mn-3d bands in a narrow energy range and thus the optical dielectric function very rapidly rises already at about 0.5 eV above the direct band gap onset. While we see an excitonic peak below the direct gap for ${\bf E}\parallel{\bf a}$ and ${\bf E}\parallel{\bf c}$  they lie above the indirect gap and for ${\bf E}\parallel{\bf b}$ no clear excitonic peak is seen. In contrast, for MnGeN$_2$, sharp excitonic peaks appear below the direct gap for all three polarizations, with the lowest one for ${\bf E}\parallel{\bf b}$. However, they still lie above the indirect gap. 
	Indirect gap excitons might also be present but cannot be calculated with our present approach because we here do not include phonon mediated optical transitions and our BSE approach only includes mixing of vertical transitions and only electronic screening. 
         Strictly speaking the only symmetry preserved in the antiferromagnetic $Pna^\prime2_1^\prime$ is the mirror plane $n_x$, so states can be even or odd with respect to this mirror plane. This is not sufficient to fully explain the polarization dependence of the spectrum as both $z$ and $y$ polarization are even and $x$ is odd with respect to this mirror plane. We find in fact that the same exciton can have both $y$ and $z$ oscillator strength and for this spin-resolved Hamiltonian, we cannot classify the bands from which they are derived in terms of the irreducible representations of the C$_{2v}$ group.              
     
     The symmetry labeling of bands near the gap along with transition energies are given in Table \ref{tablevels} and the lowest few exciton eigenvalues are given in Table \ref{tab-exciton} along with their oscillator strength relative to the first bright one, their polarization and the band of corresponding symmetry from which they are derived. Excitonic peaks seen in the BSE results in Fig. \ref{fig-eps} below the direct gap are consistent with those reported in Table \ref{tab-exciton}, however the broadening effects impede the identification of the less prominent satellite peaks. Fig. \ref{fig-exciton}(a)(b) shows the $\bf{k}$-space exciton weights, $W_{c(v)\bf k}=\sum_{v(c)}\left|A^{\lambda}_{vc\bf k}({\bf q})\right|^2$, for the first two excitons $\lambda=1,2$ which are derived from the top two valence bands to CBM in (a) MnSiN$_2$, and in (b) MnGeN$_2$.  

     We can see the lowest exciton at 3.101 eV for MnGeN$_2$ occurs for light polarization ${\bf E}\parallel {\bf b}$. This is consistent with the top of the valence band having even symmetry with respect to $n_x$ as also the conduction band minimum having even symmetry. Similarly the second exciton which occurs for polarization ${\bf E}\parallel {\bf a}$ is derived  from an odd symmetry valence band. Table \ref{tab-exciton} list for the 5 lowest excitons which band edges they are derived from and one can see that they have approximately the same binding energy and are all derived essentially from the band edges at $\Gamma$ and localized in {\bf k}-space and thus Wannier like excitons.

     The situation in MnSiN$_2$ is rather different. The lowest exciton 
     at 4.484 eV is dark and is derived from the top valence band and lowest conduction band at $\Gamma$. Because the top of the valence band at $\Gamma$ is odd and the CBM is even, it could in principle be allowed for $x$ polarization but the matrix elements turn out to negligible. The second exciton at 4.682 eV is bright with polarization along $z$ but its band decomposition shows it derives from a wide range of ${\bf k}$-points rather than from the neighborhood of the $\Gamma$- point. It is thus expected to be more localized in real space and more Frenkel type. It derives from nearly parallel regions of the top valence band and bottom conduction band, where the bands are N-$p$ like in the valence band and Mn-$d$ like bands in the conduction band. 
     The next exciton at 4.712 eV is derived from near the $\Gamma$-point but from the VBM-1 state which has even symmetry and is thus allowed for $y$ and $z$ polarization. However, it has a much smaller oscillator strength than excitonic eigenvalue 2. 
     Exciton 4  at 4.474 eV is again derived from a wide range of {\bf k}
     and has contributions from the top two valence bands. For these excitons, we cannot simply assign a single band edge and thus it is not clear what its binding energy is. For exciton 5 at 4.847 eV we return to a $\Gamma$-point derived exciton but now $x$-polarized. It is also derived from the top valence band as exciton 1 but both of these are quite weak in oscillator strength. We thus see that in MnSiN$_2$ there is an interesting mix of both Wannier type and Frenkel type charge-transfer excitons.  Note that we call these eigenvalues of the two-particle Hamiltonian excitons even though they strictly do not lie in the band gap as they are above the indirect gap. The  band weight decomposition of each of the lower 5 excitonic states are shown in Supplemental Information \cite{supinfo}.
     
     The exciton binding energies here are likely overestimated because only electronic screening is included in the BSE calculation and because a finer sampling of the Brillouin zone near $\Gamma$ may be required.
     For MnGeN$_2$, the exciton binding energy of the first 5 excitons are all about 0.26 eV, but are sensitive to the {\bf k}-point mesh. If we extrapolated from a $3\times3\times3$ mesh to a $6\times6\times6$ mesh assuming the exciton energy varies linearly with $1/N_k$ then the extrapolated value is 0.22 eV. However, in the hydrogenic model exciton binding energies are proportional to $1/\varepsilon^2$ and we really should use the static rather than electronic only dielectric constant. We have not yet calculated the static dielectric constants but in related materials this ratio is typically about 
     $(\varepsilon_\infty/\varepsilon_0)^2=0.3$. Thus we scale down our estimate of the exciton binding energy to about 0.07 eV. For MnSiN$_2$ it is less clear how to estimate the exciton binding energy because  the lowest bright exciton is not a typical Wannier type exciton. As all these excitons lie above the indirect gap one might expect them to broaden by interaction with the continuum above the indirect gap and are thus strictly excitonic resonances. However, one might expect an additional indirect exciton with similar binding energy.                 
                
        The only data on optical absorption we are aware of are from \cite{Esmaeilzadeh2006}. They measured absorption between 14500 and 30000 cm$^{-1}$ or 1.83 to 3.72 eV. They report peaks at 2.726  eV and 3.208 eV which they
        assign to internal Mn $d-d$ inter multiplet transitions.
        Near 3.7 eV they show a sharp rise of the absorption but the maximum measured is of order 2000 cm$^{-1}$. This is typical for an indirect absorption edge, while for the direct excitons we expect an $\alpha=\omega\varepsilon_2(\omega)/nc$ where $n$ is the index of refraction of order $10^4$ cm$^{-1}$.  Our $\varepsilon_1(\omega=0)$ is
        about 7, giving an index of refraction of 2.65 and with a value of $\varepsilon_2$ of about 12, we obtain an absorption strength at the lowest direct excitons of order $4\times10^4$ cm$^{-1}$. Our direct lowest excitons in MnSiN$_2$ are expected near the direct gap of 4.79 eV and thus fall outside the range measured in \cite{Esmaeilzadeh2006}. However, their onset near 3.7 eV is close to our calculated indirect gap of 3.9 eV.
        The internal $d-d$ transitions on Mn fall outside the scope of a band picture or even BSE calculation, and instead correspond to transitions between the Mn $d$ ground state $^6S$ to excited state multiplet $^4G$, which are consistent with Hund's rules. 
        
	\section{Conclusions}
	In this paper, we presented first-principles calculated electronic band structures, optical dielectric functions, magnetic exchange interactions, and spin waves for MnSiN$_2$ and MnGeN$_2$. We showed these materials belong to the class of altermagnets and verified the symmetry based predictions for spin-splitting of the bands for each of the antiferromagnetic configurations, of which we find the $Pna^\prime2_1^\prime$ spin group to have the lowest total energy. The mean field estimated N\'eel temperature is found to be in good agreement with experiment, indicating a compensation of quantum correction and beyond mean field correction factors. The materials are found to have a large indirect band gap of about 4.4 eV and 3.2 eV with lowest direct gap at $\Gamma$ of 5.1 eV and 3.5 eV  for the Si and Ge cases respectively. The spin splitting only occurs along low symmetry directions and is small near the CBM but of order 0.1 eV along $T \Gamma$ and $\Gamma R$ in the top valence band. The spin splitting is shown to manifest itself in a rotation of the constant energy surfaces between up and down spin just below the VBM and would have effects on the transport for p-type doping of these materials. 
	
\acknowledgments{This work was supported by the U.S. Department of Energy Basic Energy Sciences (DOE-BES) under Grant No. DE-SC0008933. Calculations made use of the High Performance Computing Resource in the Core Facility for Advanced Research Computing at Case Western Reserve University.}

\centerline{\bf DATA  AVAILABILITY}
The data that support the findings of this article are openly available \cite{datavail}; embargo periods may apply.

\bigskip

\bibliography{Bib/altermagnet,Bib/mn-iv-n2,Bib/lmto,Bib/dft,Bib/gw,Bib/BSE,Bib/spinchi}

\begin{thebibliography}{31}%
\makeatletter
\providecommand \@ifxundefined [1]{%
 \@ifx{#1\undefined}
}%
\providecommand \@ifnum [1]{%
 \ifnum #1\expandafter \@firstoftwo
 \else \expandafter \@secondoftwo
 \fi
}%
\providecommand \@ifx [1]{%
 \ifx #1\expandafter \@firstoftwo
 \else \expandafter \@secondoftwo
 \fi
}%
\providecommand \natexlab [1]{#1}%
\providecommand \enquote  [1]{``#1''}%
\providecommand \bibnamefont  [1]{#1}%
\providecommand \bibfnamefont [1]{#1}%
\providecommand \citenamefont [1]{#1}%
\providecommand \href@noop [0]{\@secondoftwo}%
\providecommand \href [0]{\begingroup \@sanitize@url \@href}%
\providecommand \@href[1]{\@@startlink{#1}\@@href}%
\providecommand \@@href[1]{\endgroup#1\@@endlink}%
\providecommand \@sanitize@url [0]{\catcode `\\12\catcode `\$12\catcode
  `\&12\catcode `\#12\catcode `\^12\catcode `\_12\catcode `\%12\relax}%
\providecommand \@@startlink[1]{}%
\providecommand \@@endlink[0]{}%
\providecommand \url  [0]{\begingroup\@sanitize@url \@url }%
\providecommand \@url [1]{\endgroup\@href {#1}{\urlprefix }}%
\providecommand \urlprefix  [0]{URL }%
\providecommand \Eprint [0]{\href }%
\providecommand \doibase [0]{https://doi.org/}%
\providecommand \selectlanguage [0]{\@gobble}%
\providecommand \bibinfo  [0]{\@secondoftwo}%
\providecommand \bibfield  [0]{\@secondoftwo}%
\providecommand \translation [1]{[#1]}%
\providecommand \BibitemOpen [0]{}%
\providecommand \bibitemStop [0]{}%
\providecommand \bibitemNoStop [0]{.\EOS\space}%
\providecommand \EOS [0]{\spacefactor3000\relax}%
\providecommand \BibitemShut  [1]{\csname bibitem#1\endcsname}%
\let\auto@bib@innerbib\@empty
\bibitem [{\citenamefont {Mazin}(2022)}]{Mazin2022}%
  \BibitemOpen
  \bibfield  {author} {\bibinfo {author} {\bibfnamefont {I.}~\bibnamefont
  {Mazin}} (\bibinfo {collaboration} {The PRX Editors}),\ }\href
  {https://doi.org/10.1103/PhysRevX.12.040002} {\bibfield  {journal} {\bibinfo
  {journal} {Phys. Rev. X}\ }\textbf {\bibinfo {volume} {12}},\ \bibinfo
  {pages} {040002} (\bibinfo {year} {2022})}\BibitemShut {NoStop}%
\bibitem [{\citenamefont {\ifmmode~\check{S}\else \v{S}\fi{}mejkal}\ \emph
  {et~al.}(2022{\natexlab{a}})\citenamefont {\ifmmode~\check{S}\else
  \v{S}\fi{}mejkal}, \citenamefont {Sinova},\ and\ \citenamefont
  {Jungwirth}}]{Smejkal2022}%
  \BibitemOpen
  \bibfield  {author} {\bibinfo {author} {\bibfnamefont {L.}~\bibnamefont
  {\ifmmode~\check{S}\else \v{S}\fi{}mejkal}}, \bibinfo {author} {\bibfnamefont
  {J.}~\bibnamefont {Sinova}},\ and\ \bibinfo {author} {\bibfnamefont
  {T.}~\bibnamefont {Jungwirth}},\ }\href
  {https://doi.org/10.1103/PhysRevX.12.040501} {\bibfield  {journal} {\bibinfo
  {journal} {Phys. Rev. X}\ }\textbf {\bibinfo {volume} {12}},\ \bibinfo
  {pages} {040501} (\bibinfo {year} {2022}{\natexlab{a}})}\BibitemShut
  {NoStop}%
\bibitem [{\citenamefont {\ifmmode~\check{S}\else \v{S}\fi{}mejkal}\ \emph
  {et~al.}(2022{\natexlab{b}})\citenamefont {\ifmmode~\check{S}\else
  \v{S}\fi{}mejkal}, \citenamefont {Sinova},\ and\ \citenamefont
  {Jungwirth}}]{Smejkal2022a}%
  \BibitemOpen
  \bibfield  {author} {\bibinfo {author} {\bibfnamefont {L.}~\bibnamefont
  {\ifmmode~\check{S}\else \v{S}\fi{}mejkal}}, \bibinfo {author} {\bibfnamefont
  {J.}~\bibnamefont {Sinova}},\ and\ \bibinfo {author} {\bibfnamefont
  {T.}~\bibnamefont {Jungwirth}},\ }\href
  {https://doi.org/10.1103/PhysRevX.12.031042} {\bibfield  {journal} {\bibinfo
  {journal} {Phys. Rev. X}\ }\textbf {\bibinfo {volume} {12}},\ \bibinfo
  {pages} {031042} (\bibinfo {year} {2022}{\natexlab{b}})}\BibitemShut
  {NoStop}%
\bibitem [{\citenamefont {Yuan}\ \emph {et~al.}(2020)\citenamefont {Yuan},
  \citenamefont {Wang}, \citenamefont {Luo}, \citenamefont {Rashba},\ and\
  \citenamefont {Zunger}}]{YuanLD2020}%
  \BibitemOpen
  \bibfield  {author} {\bibinfo {author} {\bibfnamefont {L.-D.}\ \bibnamefont
  {Yuan}}, \bibinfo {author} {\bibfnamefont {Z.}~\bibnamefont {Wang}}, \bibinfo
  {author} {\bibfnamefont {J.-W.}\ \bibnamefont {Luo}}, \bibinfo {author}
  {\bibfnamefont {E.~I.}\ \bibnamefont {Rashba}},\ and\ \bibinfo {author}
  {\bibfnamefont {A.}~\bibnamefont {Zunger}},\ }\href
  {https://doi.org/10.1103/PhysRevB.102.014422} {\bibfield  {journal} {\bibinfo
   {journal} {Phys. Rev. B}\ }\textbf {\bibinfo {volume} {102}},\ \bibinfo
  {pages} {014422} (\bibinfo {year} {2020})}\BibitemShut {NoStop}%
\bibitem [{\citenamefont {\ifmmode~\check{S}\else \v{S}\fi{}mejkal}\ \emph
  {et~al.}(2023)\citenamefont {\ifmmode~\check{S}\else \v{S}\fi{}mejkal},
  \citenamefont {Marmodoro}, \citenamefont {Ahn}, \citenamefont
  {Gonz\'alez-Hern\'andez}, \citenamefont {Turek}, \citenamefont {Mankovsky},
  \citenamefont {Ebert}, \citenamefont {D'Souza}, \citenamefont
  {\ifmmode~\check{S}\else \v{S}\fi{}ipr}, \citenamefont {Sinova},\ and\
  \citenamefont {Jungwirth}}]{Smejkal2023}%
  \BibitemOpen
  \bibfield  {author} {\bibinfo {author} {\bibfnamefont {L.}~\bibnamefont
  {\ifmmode~\check{S}\else \v{S}\fi{}mejkal}}, \bibinfo {author} {\bibfnamefont
  {A.}~\bibnamefont {Marmodoro}}, \bibinfo {author} {\bibfnamefont {K.-H.}\
  \bibnamefont {Ahn}}, \bibinfo {author} {\bibfnamefont {R.}~\bibnamefont
  {Gonz\'alez-Hern\'andez}}, \bibinfo {author} {\bibfnamefont {I.}~\bibnamefont
  {Turek}}, \bibinfo {author} {\bibfnamefont {S.}~\bibnamefont {Mankovsky}},
  \bibinfo {author} {\bibfnamefont {H.}~\bibnamefont {Ebert}}, \bibinfo
  {author} {\bibfnamefont {S.~W.}\ \bibnamefont {D'Souza}}, \bibinfo {author}
  {\bibfnamefont {O.~c.~v.}\ \bibnamefont {\ifmmode~\check{S}\else
  \v{S}\fi{}ipr}}, \bibinfo {author} {\bibfnamefont {J.}~\bibnamefont
  {Sinova}},\ and\ \bibinfo {author} {\bibfnamefont {T.~c.~v.}\ \bibnamefont
  {Jungwirth}},\ }\href {https://doi.org/10.1103/PhysRevLett.131.256703}
  {\bibfield  {journal} {\bibinfo  {journal} {Phys. Rev. Lett.}\ }\textbf
  {\bibinfo {volume} {131}},\ \bibinfo {pages} {256703} (\bibinfo {year}
  {2023})}\BibitemShut {NoStop}%
\bibitem [{\citenamefont {Sears}\ \emph {et~al.}(2026)\citenamefont {Sears},
  \citenamefont {Garlea}, \citenamefont {Lederman}, \citenamefont {Tranquada},\
  and\ \citenamefont {Zaliznyak}}]{Sears2026}%
  \BibitemOpen
  \bibfield  {author} {\bibinfo {author} {\bibfnamefont {J.}~\bibnamefont
  {Sears}}, \bibinfo {author} {\bibfnamefont {V.~O.}\ \bibnamefont {Garlea}},
  \bibinfo {author} {\bibfnamefont {D.}~\bibnamefont {Lederman}}, \bibinfo
  {author} {\bibfnamefont {J.~M.}\ \bibnamefont {Tranquada}},\ and\ \bibinfo
  {author} {\bibfnamefont {I.~A.}\ \bibnamefont {Zaliznyak}},\ }\href
  {https://doi.org/10.1103/g6dt-rf8c} {\bibfield  {journal} {\bibinfo
  {journal} {Phys. Rev. Lett.}\ }\textbf {\bibinfo {volume} {136}},\ \bibinfo
  {pages} {226701} (\bibinfo {year} {2026})}\BibitemShut {NoStop}%
\bibitem [{\citenamefont {Beida}\ \emph {et~al.}(2025)\citenamefont {Beida},
  \citenamefont {{\c{S}}a{\c{s}}{\i}o{\u{g}}lu}, \citenamefont {Friedrich},
  \citenamefont {Bihlmayer}, \citenamefont {Mokrousov},\ and\ \citenamefont
  {Bl{\"u}gel}}]{Beida2025}%
  \BibitemOpen
  \bibfield  {author} {\bibinfo {author} {\bibfnamefont {W.}~\bibnamefont
  {Beida}}, \bibinfo {author} {\bibfnamefont {E.}~\bibnamefont
  {{\c{S}}a{\c{s}}{\i}o{\u{g}}lu}}, \bibinfo {author} {\bibfnamefont
  {C.}~\bibnamefont {Friedrich}}, \bibinfo {author} {\bibfnamefont
  {G.}~\bibnamefont {Bihlmayer}}, \bibinfo {author} {\bibfnamefont
  {Y.}~\bibnamefont {Mokrousov}},\ and\ \bibinfo {author} {\bibfnamefont
  {S.}~\bibnamefont {Bl{\"u}gel}},\ }\href
  {https://doi.org/10.1038/s41535-025-00818-8} {\bibfield  {journal} {\bibinfo
  {journal} {npj quantum materials}\ }\textbf {\bibinfo {volume} {10}},\
  \bibinfo {pages} {97} (\bibinfo {year} {2025})}\BibitemShut {NoStop}%
\bibitem [{\citenamefont {Turan}\ \emph {et~al.}(2025)\citenamefont {Turan},
  \citenamefont {Lambrecht},\ and\ \citenamefont {Jackson}}]{Turan2025}%
  \BibitemOpen
  \bibfield  {author} {\bibinfo {author} {\bibfnamefont {I.~K.}\ \bibnamefont
  {Turan}}, \bibinfo {author} {\bibfnamefont {W.~R.~L.}\ \bibnamefont
  {Lambrecht}},\ and\ \bibinfo {author} {\bibfnamefont {J.}~\bibnamefont
  {Jackson}},\ }\href {https://doi.org/10.1103/x5w9-55dc} {\bibfield  {journal}
  {\bibinfo  {journal} {Phys. Rev. B}\ }\textbf {\bibinfo {volume} {112}},\
  \bibinfo {pages} {144443} (\bibinfo {year} {2025})}\BibitemShut {NoStop}%
\bibitem [{\citenamefont {Wintenberger}\ \emph {et~al.}(1977)\citenamefont
  {Wintenberger}, \citenamefont {Marchand},\ and\ \citenamefont
  {Maunaye}}]{Wintenberger1977}%
  \BibitemOpen
  \bibfield  {author} {\bibinfo {author} {\bibfnamefont {M.}~\bibnamefont
  {Wintenberger}}, \bibinfo {author} {\bibfnamefont {R.}~\bibnamefont
  {Marchand}},\ and\ \bibinfo {author} {\bibfnamefont {M.}~\bibnamefont
  {Maunaye}},\ }\href
  {https://doi.org/https://doi.org/10.1016/0038-1098(77)91139-5} {\bibfield
  {journal} {\bibinfo  {journal} {Solid State Communications}\ }\textbf
  {\bibinfo {volume} {21}},\ \bibinfo {pages} {733} (\bibinfo {year}
  {1977})}\BibitemShut {NoStop}%
\bibitem [{\citenamefont {Wintenberger}\ \emph {et~al.}(1972)\citenamefont
  {Wintenberger}, \citenamefont {Guyader},\ and\ \citenamefont
  {Maunaye}}]{Wintenberger1972}%
  \BibitemOpen
  \bibfield  {author} {\bibinfo {author} {\bibfnamefont {M.}~\bibnamefont
  {Wintenberger}}, \bibinfo {author} {\bibfnamefont {J.}~\bibnamefont
  {Guyader}},\ and\ \bibinfo {author} {\bibfnamefont {M.}~\bibnamefont
  {Maunaye}},\ }\href
  {https://doi.org/https://doi.org/10.1016/0038-1098(72)90505-4} {\bibfield
  {journal} {\bibinfo  {journal} {Solid State Communications}\ }\textbf
  {\bibinfo {volume} {11}},\ \bibinfo {pages} {1485} (\bibinfo {year}
  {1972})}\BibitemShut {NoStop}%
\bibitem [{\citenamefont {Kautzsch}\ \emph {et~al.}(2023)\citenamefont
  {Kautzsch}, \citenamefont {Georgescu}, \citenamefont {Puggioni},
  \citenamefont {Kent}, \citenamefont {Taddei}, \citenamefont {Reilly},
  \citenamefont {Seshadri}, \citenamefont {Rondinelli},\ and\ \citenamefont
  {Wilson}}]{Kautzsch2023}%
  \BibitemOpen
  \bibfield  {author} {\bibinfo {author} {\bibfnamefont {L.}~\bibnamefont
  {Kautzsch}}, \bibinfo {author} {\bibfnamefont {A.~B.}\ \bibnamefont
  {Georgescu}}, \bibinfo {author} {\bibfnamefont {D.}~\bibnamefont {Puggioni}},
  \bibinfo {author} {\bibfnamefont {G.}~\bibnamefont {Kent}}, \bibinfo {author}
  {\bibfnamefont {K.~M.}\ \bibnamefont {Taddei}}, \bibinfo {author}
  {\bibfnamefont {A.}~\bibnamefont {Reilly}}, \bibinfo {author} {\bibfnamefont
  {R.}~\bibnamefont {Seshadri}}, \bibinfo {author} {\bibfnamefont {J.~M.}\
  \bibnamefont {Rondinelli}},\ and\ \bibinfo {author} {\bibfnamefont {S.~D.}\
  \bibnamefont {Wilson}},\ }\href
  {https://doi.org/10.1103/PhysRevMaterials.7.104406} {\bibfield  {journal}
  {\bibinfo  {journal} {Phys. Rev. Mater.}\ }\textbf {\bibinfo {volume} {7}},\
  \bibinfo {pages} {104406} (\bibinfo {year} {2023})}\BibitemShut {NoStop}%
\bibitem [{\citenamefont {Esmaeilzadeh}\ \emph {et~al.}(2006)\citenamefont
  {Esmaeilzadeh}, \citenamefont {Hålenius},\ and\ \citenamefont
  {Valldor}}]{Esmaeilzadeh2006}%
  \BibitemOpen
  \bibfield  {author} {\bibinfo {author} {\bibfnamefont {S.}~\bibnamefont
  {Esmaeilzadeh}}, \bibinfo {author} {\bibfnamefont {U.}~\bibnamefont
  {Hålenius}},\ and\ \bibinfo {author} {\bibfnamefont {M.}~\bibnamefont
  {Valldor}},\ }\href {https://doi.org/10.1021/cm060382t} {\bibfield  {journal}
  {\bibinfo  {journal} {Chemistry of Materials}\ }\textbf {\bibinfo {volume}
  {18}},\ \bibinfo {pages} {2713} (\bibinfo {year} {2006})}\BibitemShut
  {NoStop}%
\bibitem [{\citenamefont {Lazarov}\ \emph {et~al.}(2004)\citenamefont
  {Lazarov}, \citenamefont {Li}, \citenamefont {Weinert},\ and\ \citenamefont
  {Gajdardziska-Josifovska}}]{Lazarov2004}%
  \BibitemOpen
  \bibfield  {author} {\bibinfo {author} {\bibfnamefont {V.~K.}\ \bibnamefont
  {Lazarov}}, \bibinfo {author} {\bibfnamefont {L.}~\bibnamefont {Li}},
  \bibinfo {author} {\bibfnamefont {M.}~\bibnamefont {Weinert}},\ and\ \bibinfo
  {author} {\bibfnamefont {M.}~\bibnamefont {Gajdardziska-Josifovska}},\ }\href
  {https://doi.org/10.1017/S1431927604885696} {\bibfield  {journal} {\bibinfo
  {journal} {Microscopy and Microanalysis}\ }\textbf {\bibinfo {volume} {10}},\
  \bibinfo {pages} {516} (\bibinfo {year} {2004})},\ \Eprint
  {https://arxiv.org/abs/https://academic.oup.com/mam/article-pdf/10/S02/516/48102552/mam0516.pdf}
  {https://academic.oup.com/mam/article-pdf/10/S02/516/48102552/mam0516.pdf}
  \BibitemShut {NoStop}%
\bibitem [{\citenamefont {van Schilfgaarde}\ \emph {et~al.}(2006)\citenamefont
  {van Schilfgaarde}, \citenamefont {Kotani},\ and\ \citenamefont
  {Faleev}}]{MvSQSGWprl}%
  \BibitemOpen
  \bibfield  {author} {\bibinfo {author} {\bibfnamefont {M.}~\bibnamefont {van
  Schilfgaarde}}, \bibinfo {author} {\bibfnamefont {T.}~\bibnamefont
  {Kotani}},\ and\ \bibinfo {author} {\bibfnamefont {S.}~\bibnamefont
  {Faleev}},\ }\href {https://doi.org/10.1103/PhysRevLett.96.226402} {\bibfield
   {journal} {\bibinfo  {journal} {Phys. Rev. Lett.}\ }\textbf {\bibinfo
  {volume} {96}},\ \bibinfo {pages} {226402} (\bibinfo {year}
  {2006})}\BibitemShut {NoStop}%
\bibitem [{\citenamefont {Kotani}\ \emph {et~al.}(2007)\citenamefont {Kotani},
  \citenamefont {van Schilfgaarde},\ and\ \citenamefont {Faleev}}]{Kotani2007}%
  \BibitemOpen
  \bibfield  {author} {\bibinfo {author} {\bibfnamefont {T.}~\bibnamefont
  {Kotani}}, \bibinfo {author} {\bibfnamefont {M.}~\bibnamefont {van
  Schilfgaarde}},\ and\ \bibinfo {author} {\bibfnamefont {S.~V.}\ \bibnamefont
  {Faleev}},\ }\href {https://doi.org/10.1103/PhysRevB.76.165106} {\bibfield
  {journal} {\bibinfo  {journal} {Phys.Rev. B}\ }\textbf {\bibinfo {volume}
  {76}},\ \bibinfo {eid} {165106} (\bibinfo {year} {2007})}\BibitemShut
  {NoStop}%
\bibitem [{\citenamefont {Hedin}(1965)}]{Hedin65}%
  \BibitemOpen
  \bibfield  {author} {\bibinfo {author} {\bibfnamefont {L.}~\bibnamefont
  {Hedin}},\ }\href {https://doi.org/10.1103/PhysRev.139.A796} {\bibfield
  {journal} {\bibinfo  {journal} {Phys. Rev.}\ }\textbf {\bibinfo {volume}
  {139}},\ \bibinfo {pages} {A796} (\bibinfo {year} {1965})}\BibitemShut
  {NoStop}%
\bibitem [{\citenamefont {Hedin}\ and\ \citenamefont
  {Lundqvist}(1969)}]{Hedin69}%
  \BibitemOpen
  \bibfield  {author} {\bibinfo {author} {\bibfnamefont {L.}~\bibnamefont
  {Hedin}}\ and\ \bibinfo {author} {\bibfnamefont {S.}~\bibnamefont
  {Lundqvist}},\ }in\ \href@noop {} {\emph {\bibinfo {booktitle} {Solid State
  Physics, Advanced in Research and Applications}}},\ Vol.~\bibinfo {volume}
  {23},\ \bibinfo {editor} {edited by\ \bibinfo {editor} {\bibfnamefont
  {F.}~\bibnamefont {Seitz}}, \bibinfo {editor} {\bibfnamefont
  {D.}~\bibnamefont {Turnbull}},\ and\ \bibinfo {editor} {\bibfnamefont
  {H.}~\bibnamefont {Ehrenreich}}}\ (\bibinfo  {publisher} {Academic Press},\
  \bibinfo {address} {New York},\ \bibinfo {year} {1969})\ pp.\ \bibinfo
  {pages} {1--181}\BibitemShut {NoStop}%
\bibitem [{\citenamefont {Pashov}\ \emph {et~al.}(2019)\citenamefont {Pashov},
  \citenamefont {Acharya}, \citenamefont {Lambrecht}, \citenamefont {Jackson},
  \citenamefont {Belashchenko}, \citenamefont {Chantis}, \citenamefont
  {Jamet},\ and\ \citenamefont {van Schilfgaarde}}]{questaalpaper}%
  \BibitemOpen
  \bibfield  {author} {\bibinfo {author} {\bibfnamefont {D.}~\bibnamefont
  {Pashov}}, \bibinfo {author} {\bibfnamefont {S.}~\bibnamefont {Acharya}},
  \bibinfo {author} {\bibfnamefont {W.~R.}\ \bibnamefont {Lambrecht}}, \bibinfo
  {author} {\bibfnamefont {J.}~\bibnamefont {Jackson}}, \bibinfo {author}
  {\bibfnamefont {K.~D.}\ \bibnamefont {Belashchenko}}, \bibinfo {author}
  {\bibfnamefont {A.}~\bibnamefont {Chantis}}, \bibinfo {author} {\bibfnamefont
  {F.}~\bibnamefont {Jamet}},\ and\ \bibinfo {author} {\bibfnamefont
  {M.}~\bibnamefont {van Schilfgaarde}},\ }\href
  {https://doi.org/https://doi.org/10.1016/j.cpc.2019.107065} {\bibfield
  {journal} {\bibinfo  {journal} {Computer Physics Communications}\ ,\ \bibinfo
  {pages} {107065}} (\bibinfo {year} {2019})}\BibitemShut {NoStop}%
\bibitem [{\citenamefont {Cunningham}\ \emph {et~al.}(2021)\citenamefont
  {Cunningham}, \citenamefont {Gruening}, \citenamefont {Pashov},\ and\
  \citenamefont {van Schilfgaarde}}]{Cunningham21}%
  \BibitemOpen
  \bibfield  {author} {\bibinfo {author} {\bibfnamefont {B.}~\bibnamefont
  {Cunningham}}, \bibinfo {author} {\bibfnamefont {M.}~\bibnamefont
  {Gruening}}, \bibinfo {author} {\bibfnamefont {D.}~\bibnamefont {Pashov}},\
  and\ \bibinfo {author} {\bibfnamefont {M.}~\bibnamefont {van Schilfgaarde}},\
  }\href@noop {} {\bibinfo {title} {{QS$G\hat{W}$: Quasiparticle Self
  consistent GW with ladder diagrams in W}}} (\bibinfo {year} {2021}),\ \Eprint
  {https://arxiv.org/abs/2106.05759} {arXiv:2106.05759 [cond-mat.mtrl-sci]}
  \BibitemShut {NoStop}%
\bibitem [{\citenamefont {Cunningham}\ \emph {et~al.}(2023)\citenamefont
  {Cunningham}, \citenamefont {Gr\"uning}, \citenamefont {Pashov},\ and\
  \citenamefont {van Schilfgaarde}}]{Cunningham23}%
  \BibitemOpen
  \bibfield  {author} {\bibinfo {author} {\bibfnamefont {B.}~\bibnamefont
  {Cunningham}}, \bibinfo {author} {\bibfnamefont {M.}~\bibnamefont
  {Gr\"uning}}, \bibinfo {author} {\bibfnamefont {D.}~\bibnamefont {Pashov}},\
  and\ \bibinfo {author} {\bibfnamefont {M.}~\bibnamefont {van Schilfgaarde}},\
  }\href {https://doi.org/10.1103/PhysRevB.108.165104} {\bibfield  {journal}
  {\bibinfo  {journal} {Phys. Rev. B}\ }\textbf {\bibinfo {volume} {108}},\
  \bibinfo {pages} {165104} (\bibinfo {year} {2023})}\BibitemShut {NoStop}%
\bibitem [{\citenamefont {Kotani}\ and\ \citenamefont {van
  Schilfgaarde}(2008)}]{Kotani2008}%
  \BibitemOpen
  \bibfield  {author} {\bibinfo {author} {\bibfnamefont {T.}~\bibnamefont
  {Kotani}}\ and\ \bibinfo {author} {\bibfnamefont {M.}~\bibnamefont {van
  Schilfgaarde}},\ }\href {https://doi.org/10.1088/0953-8984/20/29/295214}
  {\bibfield  {journal} {\bibinfo  {journal} {Journal of Physics: Condensed
  Matter}\ }\textbf {\bibinfo {volume} {20}},\ \bibinfo {pages} {295214}
  (\bibinfo {year} {2008})}\BibitemShut {NoStop}%
\bibitem [{dat()}]{datavail}%
  \BibitemOpen
  \href@noop {} {}\bibinfo {note}
  {\url{https://github.com/Electronic-Structure-Group/Mn-IV-N2}}\BibitemShut
  {NoStop}%
\bibitem [{\citenamefont {Onida}\ \emph {et~al.}(2002)\citenamefont {Onida},
  \citenamefont {Reining},\ and\ \citenamefont {Rubio}}]{Onida02}%
  \BibitemOpen
  \bibfield  {author} {\bibinfo {author} {\bibfnamefont {G.}~\bibnamefont
  {Onida}}, \bibinfo {author} {\bibfnamefont {L.}~\bibnamefont {Reining}},\
  and\ \bibinfo {author} {\bibfnamefont {A.}~\bibnamefont {Rubio}},\ }\href
  {https://doi.org/10.1103/RevModPhys.74.601} {\bibfield  {journal} {\bibinfo
  {journal} {Rev. Mod. Phys.}\ }\textbf {\bibinfo {volume} {74}},\ \bibinfo
  {pages} {601} (\bibinfo {year} {2002})}\BibitemShut {NoStop}%
\bibitem [{\citenamefont {Strinati}(1988)}]{Strinati1988}%
  \BibitemOpen
  \bibfield  {author} {\bibinfo {author} {\bibfnamefont {G.}~\bibnamefont
  {Strinati}},\ }\href {https://doi.org/10.1007/BF02725962} {\bibfield
  {journal} {\bibinfo  {journal} {La Rivista del Nuovo Cimento (1978-1999)}\
  }\textbf {\bibinfo {volume} {11}},\ \bibinfo {pages} {1} (\bibinfo {year}
  {1988})}\BibitemShut {NoStop}%
\bibitem [{\citenamefont {Hanke}(1978)}]{Hanke78}%
  \BibitemOpen
  \bibfield  {author} {\bibinfo {author} {\bibfnamefont {W.}~\bibnamefont
  {Hanke}},\ }\href {https://doi.org/10.1080/00018737800101384} {\bibfield
  {journal} {\bibinfo  {journal} {Advances in Physics}\ }\textbf {\bibinfo
  {volume} {27}},\ \bibinfo {pages} {287} (\bibinfo {year} {1978})}\BibitemShut
  {NoStop}%
\bibitem [{\citenamefont {Rohlfing}\ and\ \citenamefont
  {Louie}(2000)}]{Rohlfing2000}%
  \BibitemOpen
  \bibfield  {author} {\bibinfo {author} {\bibfnamefont {M.}~\bibnamefont
  {Rohlfing}}\ and\ \bibinfo {author} {\bibfnamefont {S.~G.}\ \bibnamefont
  {Louie}},\ }\href {https://doi.org/10.1103/PhysRevB.62.4927} {\bibfield
  {journal} {\bibinfo  {journal} {Phys. Rev. B}\ }\textbf {\bibinfo {volume}
  {62}},\ \bibinfo {pages} {4927} (\bibinfo {year} {2000})}\BibitemShut
  {NoStop}%
\bibitem [{\citenamefont {Tyablikov}(1959)}]{Tyablikov1959}%
  \BibitemOpen
  \bibfield  {author} {\bibinfo {author} {\bibfnamefont {S.~V.}\ \bibnamefont
  {Tyablikov}},\ }\href@noop {} {\bibfield  {journal} {\bibinfo  {journal}
  {Ukr. Mat. Zh.}\ }\textbf {\bibinfo {volume} {11}},\ \bibinfo {pages} {287}
  (\bibinfo {year} {1959})}\BibitemShut {NoStop}%
\bibitem [{\citenamefont {Callen}(1963)}]{Callen63}%
  \BibitemOpen
  \bibfield  {author} {\bibinfo {author} {\bibfnamefont {H.~B.}\ \bibnamefont
  {Callen}},\ }\href {https://doi.org/10.1103/PhysRev.130.890} {\bibfield
  {journal} {\bibinfo  {journal} {Phys. Rev.}\ }\textbf {\bibinfo {volume}
  {130}},\ \bibinfo {pages} {890} (\bibinfo {year} {1963})}\BibitemShut
  {NoStop}%
\bibitem [{\citenamefont {Rusz}\ \emph {et~al.}(2005)\citenamefont {Rusz},
  \citenamefont {Turek},\ and\ \citenamefont {Divi\ifmmode~\check{s}\else
  \v{s}\fi{}}}]{Rusz2005}%
  \BibitemOpen
  \bibfield  {author} {\bibinfo {author} {\bibfnamefont {J.}~\bibnamefont
  {Rusz}}, \bibinfo {author} {\bibfnamefont {I.}~\bibnamefont {Turek}},\ and\
  \bibinfo {author} {\bibfnamefont {M.}~\bibnamefont
  {Divi\ifmmode~\check{s}\else \v{s}\fi{}}},\ }\href
  {https://doi.org/10.1103/PhysRevB.71.174408} {\bibfield  {journal} {\bibinfo
  {journal} {Phys. Rev. B}\ }\textbf {\bibinfo {volume} {71}},\ \bibinfo
  {pages} {174408} (\bibinfo {year} {2005})}\BibitemShut {NoStop}%
\bibitem [{\citenamefont {Hass}\ \emph {et~al.}(1984)\citenamefont {Hass},
  \citenamefont {Velick\'y},\ and\ \citenamefont {Ehrenreich}}]{Hass1984}%
  \BibitemOpen
  \bibfield  {author} {\bibinfo {author} {\bibfnamefont {K.~C.}\ \bibnamefont
  {Hass}}, \bibinfo {author} {\bibfnamefont {B.}~\bibnamefont {Velick\'y}},\
  and\ \bibinfo {author} {\bibfnamefont {H.}~\bibnamefont {Ehrenreich}},\
  }\href {https://doi.org/10.1103/PhysRevB.29.3697} {\bibfield  {journal}
  {\bibinfo  {journal} {Phys. Rev. B}\ }\textbf {\bibinfo {volume} {29}},\
  \bibinfo {pages} {3697} (\bibinfo {year} {1984})}\BibitemShut {NoStop}%
\bibitem [{sup()}]{supinfo}%
  \BibitemOpen
  \href@noop {} {}\bibinfo {note} {Supplemental information is available at the
  website of this publication, and gives band weights for the excitons similar
  to Fig. \ref{fig-exciton} for the other excitons listed in
  Table\ref{tab-exciton}}\BibitemShut {NoStop}%
\end{thebibliography}%

\appendix
\section{Supplemental Material}
\begin{figure*}
	\includegraphics[width=18cm]{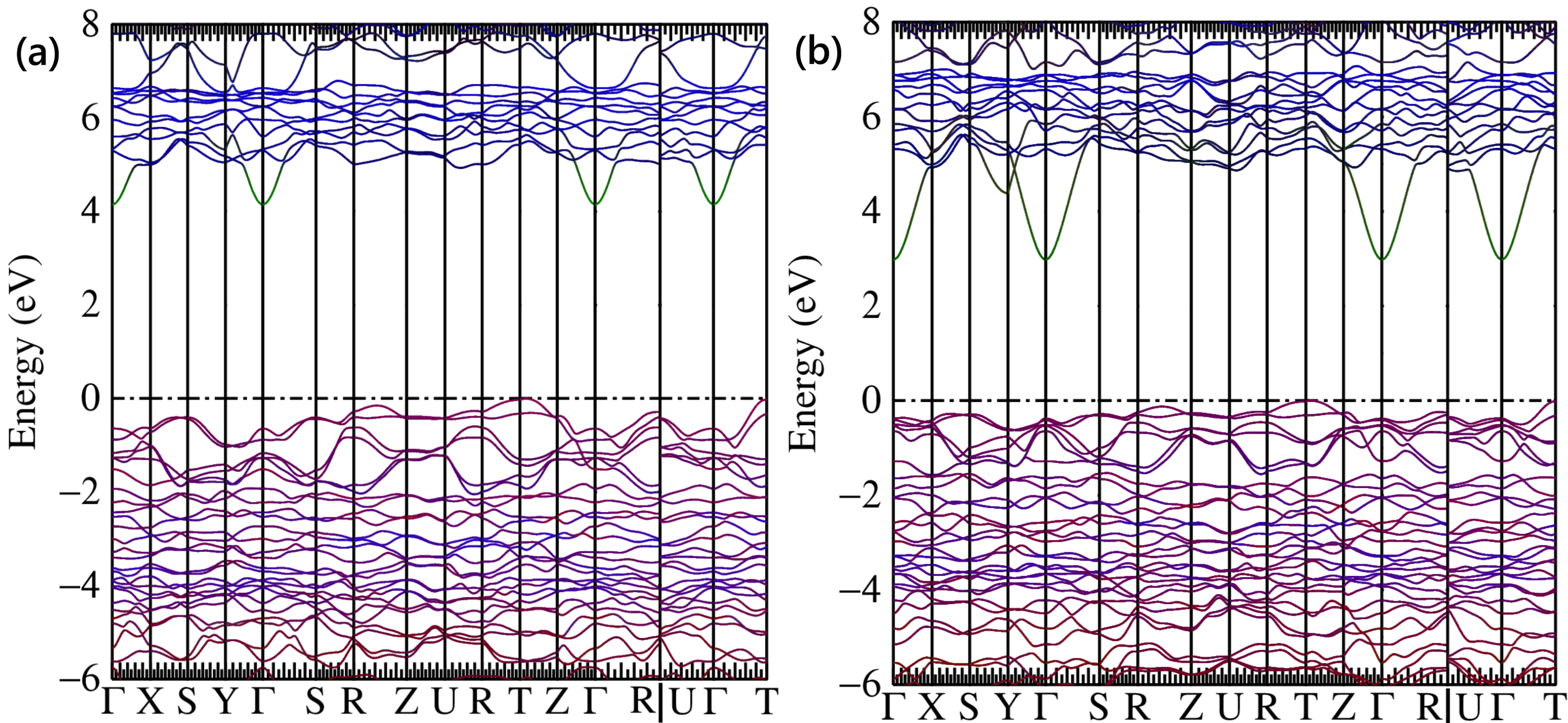}
	\caption{Majority spin bands colored with respect to their orbital characters in (a) MnSiN$_2$ and (b) MnGeN$_2$. Mn-$d$ orbitals are shown in blue, Si(Ge)-$s$ orbitals are shown green, N-$p$ orbitals are shown in red.}
	\label{fig-colband}
\end{figure*}

\begin{figure*}
	\includegraphics[width=18cm]{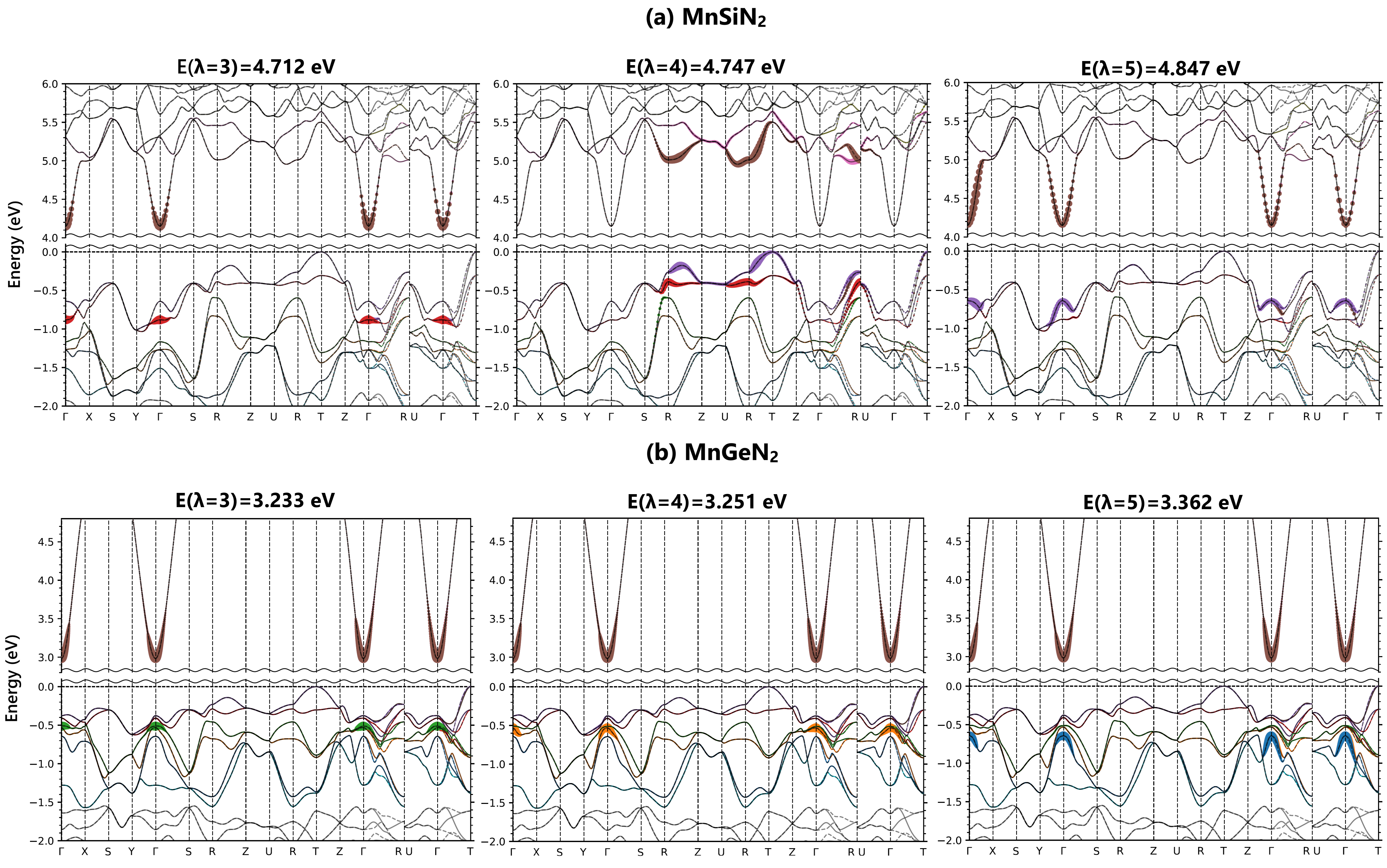}
	\caption{Exciton wave function weights $W_{v(c)\bf k}^{\lambda=3,4,5}$ in (a) MnSiN$_2$, and in (b) MnGeN$_2$. 
		\label{fig-exciton345}}
\end{figure*}

Fig. \ref{fig-colband} shows an orbital character weighted band structure for $Pna^\prime2_1^\prime$ MnSiN$_2$, showing that the top of the valence band is N-$p$-like, the conduction band minimum is Si-$s$ like at $\Gamma$ but Mn-$d$ like over most of the Brillouin zone.
Fig. \ref{fig-exciton345}(a) show the exciton band weights for excitons 3,4,5 in MnSiN$_2$ as defined in the Table VII in the main paper and Fig. \ref{fig-exciton345}(b) shows the exciton band weights for excitons 3,4,5 in MnGeN$_2$.

\end{document}